\documentclass[sigconf]{acmart}

\AtBeginDocument{%
  \providecommand\BibTeX{{%
    \normalfont B\kern-0.5em{\scshape i\kern-0.25em b}\kern-0.8em\TeX}}}

\usepackage{pgfgantt}
\usepackage{tikz}
\usetikzlibrary{shapes,arrows,positioning}
\usepackage{xurl}
\usepackage{hyperref}

\copyrightyear{2026}
\acmYear{2026}
\setcopyright{cc}
\setcctype{by}
\acmConference[CHI '26]{Proceedings of the 2026 CHI Conference on Human Factors in Computing Systems}{April 13--17, 2026}{Barcelona, Spain}
\acmBooktitle{Proceedings of the 2026 CHI Conference on Human Factors in Computing Systems (CHI '26), April 13--17, 2026, Barcelona, Spain}
\acmPrice{}
\acmDOI{10.1145/3772318.3791272}
\acmISBN{979-8-4007-2278-3/2026/04}

\usepackage{etoolbox,lipsum}
\begin{document}

\title[Experience-Based Co-Design of a Multi-modal 3D Data Visualization Tool]{Three Modalities, Two Design Probes, One Prototype, and No Vision: Experience-Based Co-Design of a Multi-modal 3D Data Visualization Tool}

\begin{teaserfigure}
    \centering
    \includegraphics[width=0.7\linewidth]{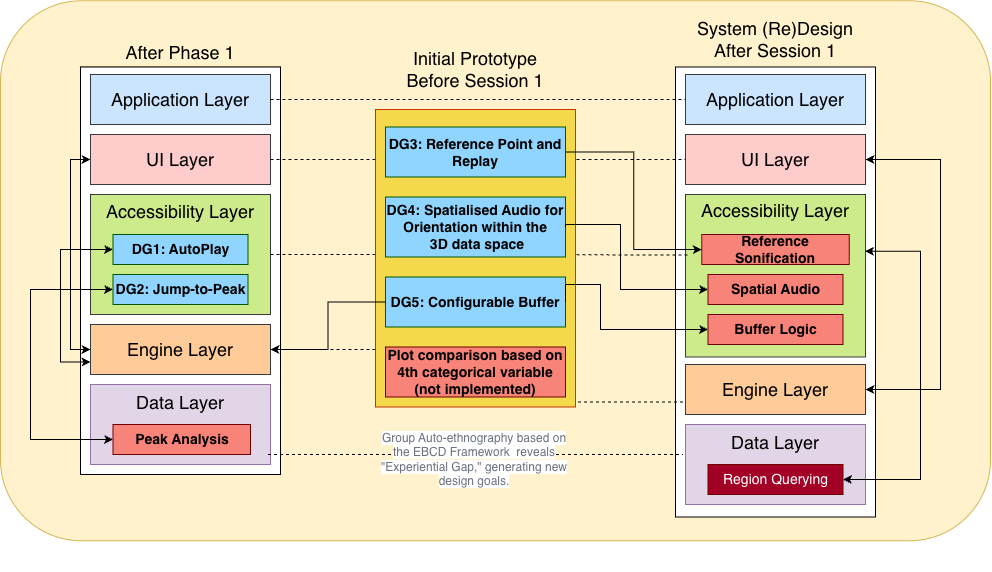}
    \caption{Iterative Prototyping based on the EBCD Framework}
    \Description{Process diagram showing a three-stage EBCD iteration. Left: Initial system with five-layer architecture including AutoPlay and Jump-to-Peak features in the Accessibility Layer. Center: Co-design session reveals experiential gap, generating new design goals for reference points, spatial audio, buffers, and plot comparison. Right: Redesigned system with enhanced Accessibility Layer featuring Reference Sonification, Spatial Audio, and Buffer Logic, plus Data Layer improvements. Arrows show how user feedback directly shaped feature development through iterative co-design.}
    \label{fig:teaser}
\end{teaserfigure}

\author{Sanchita S. Kamath}
\affiliation{%
 \department{School of Information Sciences}
  \institution{University of Illinois Urbana-Champaign}
  \orcid{0000-0001-6469-0360}
  \city{Champaign}
  \state{Illinois}
  \country{USA}
  \postcode{61820}
  }
\email{ssk11@illinois.edu}

\author{Aziz Zeidieh}
\orcid{0009-0000-9334-8660}
\affiliation{%
 \department{Informatics}
  \institution{University of Illinois Urbana-Champaign}
  \city{Champaign}
  \state{Illinois}
  \country{USA}
  \postcode{61820}
  }
\email{azeidi2@illinois.edu}

\author{Venkatesh Potluri}
\orcid{0000-0002-5027-8831}
\affiliation{%
 \department{School of Information}
  \institution{University of Michigan}
  \city{Ann-Harbour}
  \state{Michigan}
  \country{USA}
  \postcode{48109}
  }
\email{potluriv@umich.edu}

\author{Sile O'Modhrain}
\orcid{0000-0003-3804-5469}
\affiliation{%
 \department{School of Information}
  \institution{University of Michigan}
  \city{Ann-Harbour}
  \state{Michigan}
  \country{USA}
  \postcode{48109}
  }
\email{sileo@umich.edu}

\author{Kenneth Perry}
\orcid{0009-0004-0412-0373}
\affiliation{%
 \department{}
  \institution{American Printing House for the Blind}
  \city{Louisville}
  \state{Kentucky}
  \country{USA}
  \postcode{40219}
  }
\email{kperry@blinksoft.com}

\author{JooYoung Seo}
\orcid{0000-0002-4064-6012}
\affiliation{%
 \department{School of Information Sciences}
  \institution{University of Illinois Urbana-Champaign}
  \city{Champaign}
  \state{Illinois}
  \country{USA}
  \postcode{61820}
  }
\email{jseo1005@illinois.edu}

\renewcommand{\shortauthors}{Kamath et al.}

\begin{abstract}
Three-dimensional (3D) data visualizations, such as surface plots, are vital in STEM fields from biomedical imaging to meteorology and spectroscopy, yet remain largely inaccessible to blind and low-vision (BLV) people. To address this gap, we conducted an Experience-Based Co-Design (EBCD) with BLV co-designers with expertise in non-visual data representations to create an accessible, multi-modal, web-native visualization tool. Using a multi-phase co-design methodology, our team of five BLV and one non-BLV researcher(s) participated in two iterative sessions, comparing a low-fidelity tactile probe with a high-fidelity digital prototype. This process produced a prototype with empirically grounded features, including reference sonification, stereo and volumetric audio, and configurable buffer aggregation, which our BLV co-designers validated as improving analytic accuracy and learnability. In this study, we explicitly target core analytic tasks essential for non-visual 3D data exploration: 3D orientation, landmark and peak finding, comparing local maxima versus global trends, gradient tracing, and identifying occluded or partially hidden features. Our work offers accessibility researchers and developers a co-design protocol for translating tactile knowledge to digital interfaces, concrete design guidance for future systems, and opportunities to extend accessible 3D visualization into embodied data environments.

Three-dimensional data visualizations, such as surface plots, are vital in STEM fields from biomedical imaging to meteorology and spectroscopy, yet remain largely inaccessible to blind and low-vision people. To address this gap, we conducted an Experience-Based Co-Design with BLV co-designers with expertise in non-visual data representations to create an accessible, multi-modal, web-native visualization tool. Using a multi-phase co-design methodology, our team of five BLV and one non-BLV researcher(s) participated in two iterative sessions, comparing a low-fidelity tactile probe with a high-fidelity digital prototype. This process produced a prototype with empirically grounded features, including reference sonification, stereo and volumetric audio, and configurable buffer aggregation, which our co-designers validated as improving analytic accuracy and learnability. In this study, we explicitly target core analytic tasks essential for non-visual 3D data exploration: 3D orientation, landmark and peak finding, comparing local maxima versus global trends, gradient tracing, and identifying occluded or partially hidden features. Our work offers accessibility researchers and developers a co-design protocol for translating tactile knowledge to digital interfaces, concrete design guidance for future systems, and opportunities to extend accessible 3D visualization into embodied data environments.
\end{abstract}




\maketitle

\section{Introduction}
\label{sec:introduction}

\begin{figure}
    \centering
    \includegraphics[width=\linewidth]{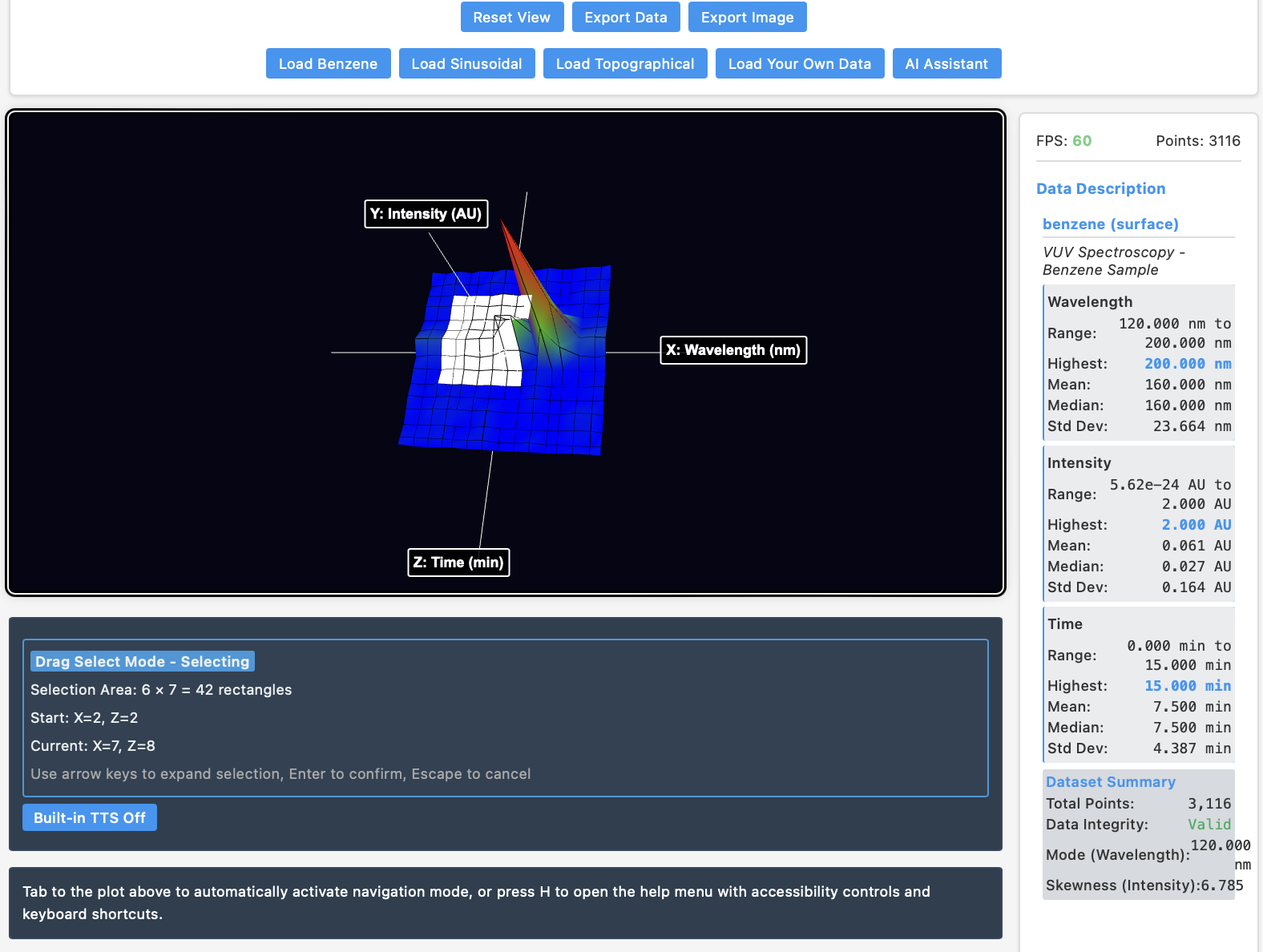}
    \caption{Prototype interface}
    \label{fig:interface}
    \Description{The interface figure depicts our browser-based, multi-modal 3D data visualization tool presenting a benzene spectroscopy dataset. At the center is a black-background panel containing a 3D surface plot with three clearly labeled axes: the X-axis represents wavelength in nanometers, the Y-axis represents intensity in arbitrary units (AU), and the Z-axis represents time in minutes. The surface is shown as a mesh grid with color gradients ranging from blue in lower intensity regions, through green and yellow, to red at the sharpest peak, visually emphasizing the area of strongest intensity. To the right of the plot, a sidebar provides dataset statistics in textual form. Under “Wavelength,” the values range from 120 to 200 nanometers, with the highest value at 200 nm, a mean of 160 nm, and a standard deviation of 23.664 nm. Under “Intensity,” the range spans from near zero to 2 AU, with a maximum of 2 AU, a mean of 0.061 AU, and a median of 0.027 AU. Under “Time,” the dataset ranges from 0 to 15 minutes, with the longest duration at 15 minutes, a mean of 7.5 minutes, and a standard deviation of 4.387 minutes. The dataset summary at the bottom of the sidebar reports a total of 3,116 data points, notes that data integrity is valid, identifies the mode wavelength as 120 nm, and lists the skewness of intensity as 6.785. Below the visualization, a blue navigation panel labeled “Drag Select Mode – Selecting” describes the current user interaction. It specifies a rectangular selection area of 6 by 7 grid units, notes the starting position (X=2, Z=2) and the current position (X=7, Z=8), and provides instructions for navigating with arrow keys, confirming with the Enter key, or canceling with the Escape key. A separate toggle button allows enabling or disabling built-in text-to-speech output. At the very top of the interface, a row of blue action buttons offers options to reset the view, export data, or export the current image, along with dataset loading options such as “Load Benzene,” “Load Sinusoidal,” and “Load Your Own Data,” as well as an “AI Assistant” button. Collectively, the figure shows how the tool integrates a 3D surface rendering, detailed numerical summaries, and interactive controls to enable exploration and analysis of spectroscopy data.}
\end{figure}

While there have been notable advances in producing accessible data visualizations~\cite{ahmetovicAudioFunctionswebMultimodalExploration2019, adams_blooms_2015, abdolrahmani_blind_2020}, blind and low-vision (BLV) people remain largely excluded from engaging with three-dimensional (3D) data visualizations that represent data using three axes (X, Y, and Z) to create spatial models, such as 3D scatter plots or volumetric rendering. Most accessibility efforts to date, have focused on audiotactile graphs ~\cite{ahmetovicAudioFunctionswebMultimodalExploration2019, huang_auditory_2012} and two-dimensional representations~\cite{butler_technology_2021}. The lack of accessible 3D data visualization tools is particularly pressing given that 3D visualizations are central to fields such as meteorology~\cite{rautenhaus_three-dimensional_2015}, biomedical imaging~\cite{zhou_review_2022}, VUV spectroscopy~\cite{kaplitz_gas_2023}, geoscience~\cite{qi_3d_2007}, and computational fluid dynamics~\cite{harte_second_2024}; when inaccessible, they not only limit analytic rigor but also alienate BLV individuals from meaningful participation in knowledge production across these domains. Our target data type in this work is continuous height-field surfaces—scalar fields defined over a two-dimensional domain that produce smooth, continuous 3D cloth-like surfaces commonly found in scientific surface plots such as in VUV spectroscopy viewable in Figure~\ref{fig:interface}. Focusing on these continuous surfaces allows us to support analytic tasks that depend on stable topological continuity.
  
While accessibility for 2D charts has progressed through tactile graphics~\cite{moured_accessible_2023, dzhurynskyi_enhancing_2024, khan_tactilenet_2025} and multi-modal, web-based interactions~\cite{zhang_charta11y_2024, hoque_accessible_2023, seo_maidr_2024}, the spatial and complex nature of 3D data representations such as surface and point plots continue to pose persistent barriers to non-visual access. Yet without access to these visualizations, analysts risk missing essential insights in spatially complex datasets~\cite{bui_role_2021, eilola_3d_2023}. 
The leap from two-dimensional (2D) charts to interactive three-dimensional (3D) data reveals critical gaps in current accessibility research. First, most studies still concentrate on 2D charts, leaving a gap in methods for interactively exploring 3D surfaces and point clouds~\cite{joyner_visualization_2022, kamath_explore_2025}. It is also important to clarify the scope of this claim within established visualization design practice, which generally cautions against projecting three-dimensional data onto two-dimensional displays due to perceptual distortions and the potential for misleading interpretations. In many cases, two-dimensional encodings such as color gradients, glyph size, or textual summaries can successfully convey three-variable correlations, and users are often able to infer relationships from such encodings. However, when analytic tasks rely on understanding continuous surface structure or spatial topology, as is common with heightfield-style representations that remain prevalent in practice, collapsing a three-dimensional topology into a flat, single-channel representation can remove critical structural relationships. For BLV users in particular, TTS-only descriptions or reduced two-dimensional encodings may force the reconstruction of three-dimensional spatial logic mentally across modalities, increasing cognitive load and weakening the formation of coherent spatial mental models. Rather than asserting that three-dimensional representations are universally superior, this work focuses on accessibility challenges that arise in contexts where three-dimensional surface representations are already in use and where preserving topological structure is central to the analytic task.

Second, to the best of our knowledge, no studies have applied evidence-based methodologies for transferring the spatial knowledge gained from tactile prototypes into web-based digital environments for the specific challenge of 3D plots~\cite{snider_mixed_2024, speicher_designers_2021}. Third, prior systems often neglect the pedagogical scaffolding required for learners’ transition from initial orientation to independent analytic reasoning~\cite{martinez-maldonado_latux_2015}. Fourth, few studies provide fine-grained evidence on how specific non-visual modalities affect performance in 3D data visualization contexts; existing work is often limited to object recognition applications~\cite{rojas_effect_2013, oliver_stereo_2018} or focuses on embodied interaction without exploring multi-modality in depth~\cite{ball_realizing_2007}. 

This paper introduces a methodology (inspired by the Experience-Based Co-Design (EBCD) framework~\cite{raynor_experiencebased_2020}) and two design probes which were used to prompt the non-visual exploration of 3D scientific data. Our work was motivated by a foundational goal: to ensure that accessibility tools are developed with and by the BLV community, not just for them. Instead of positioning BLV individuals as participants in user studies framed by sighted researchers, rather than as integral collaborators from a project's inception~\cite{AccessibilityResearchMack2018}; this study aims to flip that order. By adopting an EBCD approach~\cite{raynor_experiencebased_2020}, our work while proposed by a sighted researcher, is fundamentally designed with and inspired by BLV people. Unlike conventional co-design, which often focuses on collaborative ideation and prototyping, EBCD emphasizes grounding design in lived experiences, typically through in-depth narrative accounts, storytelling, and shared reflection. This distinction was crucial for our project: rather than relying solely on structured design workshops, we began by gathering experiential narratives from BLV co-designers who have expert knowledge in data visualization, which then directly informed the framing of our design challenges and priorities. In this way, EBCD shaped not just how we collaborated, but also what we considered meaningful design outcomes. The methodology ensured that our technical explorations were continually anchored in co-designers’ everyday practices and values, making the resulting tools not only functional but also genuinely usable and empowering. Within this process, our produced tactile probe (Section~\ref{subsec:lowfi_design-probe}) served as the shared ground-truth reference against which all subsequent digital translation was calibrated, enabling consistent interpretation of spatial features across modalities.
The significance of this work, therefore, is threefold: (1) it addresses the challenge of producing a transferable co-design protocol for translating tactile knowledge into effective digital interactions, (2) demonstrates a collaborative and inclusive research methodology - by adopting the EBCD Framework for HCI research on non-visual data access and representation, and (3) provides an evidence-based model for developing effective non-visual tools for 3D data exploration. By making surface visualizations accessible, we expand opportunities for BLV users to engage directly with spatial data, promoting fuller participation in STEM disciplines~\cite{jiang_designing_2024}. Our developed prototype, as evaluated by our ethnographic approach with expert data visualization researchers, helps demonstrate a successful translation of embodied, tactile knowledge into a well-known and well-adopted digital medium, providing non-visual features that were validated by our expert collaborators as improving their ability to orient, analyze, and accurately interpret 3D data. This gives our system a better position to help BLV people in producing spatial mental models of three-dimensional data, as compared to two-dimensional representations of the same data. The uniqueness of this work lies in demonstrating how one can transform a tactile learning experience into an effective digital counterpart. Such an approach equips educators with concrete workflows for teaching 3D spatial reasoning and provides developers with empirically grounded guidance for extending access to embodied or three-dimensional environments. While our system can handle the input of closed-volume or molecular surface data, those structures currently appear as disjoint surfaces due to our coordinate grid handling algorithm, and thus do not behave equivalently; producing non-intuitive interaction, making any transfer only partial and not assumed. Hence, we shall not focus on them in this paper.

The research problem at hand, therefore, is to make 3D data visualizations accessible and learnable for BLV people. Following are the research questions that direct this study:
\begin{itemize}
    \item [\textbf{RQ1.}] How can spatial knowledge gained through tactile, low-fidelity prototypes be effectively transferred into high-fidelity digital experiences that support core 3D analytic tasks? 
    \item [\textbf{RQ2.}] How can we (re)design specific non-visual features to empower BLV users’ ability to orient, analyze and accurately interpret 3D surfaces with confidence?
    \item [\textbf{RQ3.}] What mediums and interaction tools can most effectively translate a web-based digital prototype of three-dimensional data visualizations into embodied forms of engagement that preserve users’ ability to execute the full set of analytic tasks?
\end{itemize}

This scholarly work makes four key contributions - (1) it offers a co-design protocol that integrates tactile and digital probes, refined iteratively across two sessions with BLV collaborators; (2) it delivers a (re)designed high-fidelity prototype incorporating empirically grounded features such as reference sonification, stereo and volumetric audio cues, and configurable buffer aggregation; (3) it provides expert opinion on the impact of these features on analytic accuracy and user learnability in 3D data contexts; and (4) it contributes design ideas for the future extension to embodied 3D data visualization environments.
\section{Related Work}
\label{sec:related_work}

To contextualize our contribution, we review the EBCD theoretical framework and map how it can be employed in HCI practices and list three areas of related work: adaptations of two-dimensional visualizations for BLV audiences, immersive and embodied approaches to data interaction, and emerging efforts in three-dimensional visualization. Together, these streams highlight both the progress made and the critical gaps that remain, underscoring the need for methods that can center lived experience in tackling new accessibility challenges. 

\subsection{Theoretical Framework: Experience-Based Co-Design}
\label{subsec:ebcd-mapping}
EBCD, originally developed in health services, is built upon iterative cycles of understanding lived experience, prioritizing issues, and (re)designing services with users and stakeholders \cite{morley_evidence-informed_2024, fylan_using_2021, green_use_2020}. According to EBCD practitioners, key stages include gathering staff and patient experiences, producing a ``trigger film'' based on emotional touch-points, holding co-design events, forming working groups, and celebrating progress \cite{graber_reflections_2024}. Although participatory and co-design approaches are common in HCI \cite{sanders_co-creation_2008}, EBCD is distinguished by its systematic elevation of lived experience as central design material; not merely an input but the motivating, framing resource \cite{marwaa_using_2023, francis-auton_exploring_2024}. To the best of our knowledge, no prior HCI studies have employed EBCD for accessible 3D data visualization; we adopt and adapt it (Table~\ref{tab:ebcd-adaptation}) to structure cycles of eliciting BLV collaborators’ tactile and narrative experiences, prioritizing their analytical challenges, and collaboratively redesigning non-visual 3D representations in a meaningful, grounded way.  

\begin{table*}[t]
  \centering
  \scriptsize
  \begin{tabular}{p{4cm} p{13cm}}
    \toprule
    \textbf{Canonical EBCD Stage} & \textbf{Our Adaptation and Rationale} \\
    \midrule
    Setting up the project & We established our co-design team, clarified roles, and oriented BLV researchers to both the tactile and digital phases, ensuring a shared understanding and equal power in design. \\
    Gather staff / practitioner experiences & In our context, “staff” maps to non-BLV researcher; we collected their perspectives on constraints and design possibilities, balancing with BLV lived experience. \\
    Gather patient / user experiences & We ran low-fidelity and high-fidelity design probe sessions and narrative interviews to collect first-person accounts of exploring 3D surfaces non-visually. \\
    Trigger film (or tangible trigger) & Instead of a video film, we used the tactile probe and semi-structured questioning to surface emotional and perceptual “touch-points” in how BLV collaborators explore surface features. \\
    First co-design event & We brought BLV researchers together in our workshops, using prior experiential data to align on priorities and co-create interaction features through Session 1 and 2. \\
    Co-design working groups & We iteratively developed our high-fidelity prototype (drawing from insights gathered during Session 1) and tested the final version with requested features (e.g., sonification and buffer aggregation) during Session 2. \\
    Celebration / review event & We reserved the second half of Session 2 for reflection, where participants compared tactile and digital experiences, surfaced insights, and affirmed shared design ownership. \\
    Implementation & We integrated the co-designed features into a web-native 3D visualization tool usable for 3D data analysis by BLV users. \\
    Evaluation & We performed iterative evaluation across both sessions where BLV collaborators used both the tactile and web-based versions, reflecting on accuracy, learnability, and fidelity. \\
    \bottomrule
  \end{tabular}
  \caption{Mapping from canonical EBCD stages (per standard health-services literature) to our adaptations in this work.}
  \label{tab:ebcd-adaptation}
  \Description{Table mapping nine canonical EBCD stages from healthcare literature to HCI adaptations. Each row shows how traditional stages (e.g., gather patient experiences, trigger film, co-design events) were translated for accessible 3D visualization design with BLV researchers, including tactile probes, design sessions, iterative development, and evaluation cycles.}
\end{table*}

Compared to design studies, which often emphasize artifact critique, visual aesthetics, and designer reflection ~\cite{huang_experiential_2023}, and more conventional co-design methods, which may solicit stakeholder input during ideation~\cite{jones_contexts_2018}, EBCD offers a distinct advantage: it treats lived experience as the core design material ~\cite{ebcdMorley2024}. In design studies, the focus may lie on how form and function evolve through prototyping and critique, but without grounding in deep personal narrative ~\cite{grimaldi_narratives_2013}. In typical co-design, participants might guide requirements or suggest features, but their stories are often abstracted ~\cite{obrien_integrating_2016}. EBCD, by contrast, privileges participants’ emotional ``touch-points'': their real, embodied encounters, and uses these to prioritize problems, spark design ideas (e.g., via trigger films/probes), and evaluate solutions. For our work, this is especially powerful: the way BLV individuals navigate and make sense of tactile spatial data is not just a usability issue but a form of domain knowledge. By centering that experience, we ensure the resulting design is not only functional, but deeply aligned with how BLV users perceive, reason about, and interpret 3D data — leading to a more meaningful, effective, and empowering tool.

\subsection{Accessible Visualization in Two Dimensions}
\label{subsec:accessible2D}
Accessibility research has historically concentrated on two- dimensional charts, tactile graphics, and physical data representations. Research on haptic data visualization highlights how touch-based interfaces can support data access for BLV individuals by translating charts, maps, and diagrams into tactile forms~\cite{he_using_2025, butler_technology_2021, dzhurynskyi_enhancing_2024, khan_tactilenet_2025, moured_accessible_2023}. Physical visualization and data sculptures have been proposed as educational tools to represent abstract data tangibly, enabling learners to grasp complex relationships through embodied interaction~\cite{vandemoerePhysicalVisualizationInformation2010}. Data physicalizations have been shown to support effective analysis of elevation data when compared with digital or VR alternatives, emphasizing the promise of haptic perception in learning spatial structures~\cite{hermanTouchingGroundEvaluating2025}.

Complementing tactile approaches, sonification-based systems have demonstrated effectiveness in conveying data through auditory mappings. ~\citet{zhao_embodiment_2008} pioneered interactive sonification techniques for georeferenced data, while more recent work has integrated natural sound representations~\cite{hoque_accessible_2023} and spatial audio cues~\cite{siu_virtual_2020} to enhance data comprehension. Electrotactile feedback systems~\cite{jiang_designing_2024} and refreshable tactile displays combined with conversational agents~\cite{reinders_when_2025, holloway_animations_2022, holloway_infosonics_2022} have expanded multimodal possibilities for chart comprehension. Haptic techniques such as ~\citet{fan_slide-tone_2022} and broader sonification frameworks~\cite{ali_sonify_2020} have further enriched non-visual data access strategies.
Screen-reader-based approaches, including tools like VoxLens~\cite{sharif_voxlens_2022}, MAIDR~\cite{seo_maidr_2024}, ChartA11y~\cite{zhang_charta11y_2024}, AudioFunctions.Web~\cite{ahmetovic_audiofunctionsweb_2019}, and Umwelt~\cite{zong_umwelt_2024}, have established frameworks for keyboard-driven navigation and semantic description of visualizations. Studies of screen-reader users' experiences~\cite{sharif_understanding_2021} and rich screen-reader interaction designs~\cite{zong_rich_2022} have illuminated the diverse needs and preferences of BLV individuals when accessing visual data.

Broader surveys have documented the state of sonification -visualization integration~\cite{enge_open_2024}, established design frameworks for accessible visualization~\cite{marriott_inclusive_2021, elavsky_how_2022}, and articulated the need for sociotechnical perspectives that center disability justice~\cite{lundgard_sociotechnical_2019, hsueh_cripping_2023}. Work on semantic levels of natural language description~\cite{lundgard_accessible_2021} and identifying opportunities for accessible visualization~\cite{kim_accessible_2021} has further shaped our understanding of how to design for diverse sensory experiences. Additional research has explored sensory substitution~\cite{chundury_towards_2022}, non-visual visualization design~\cite{choi_visualizing_2019}, audio-based infographics~\cite{holloway_infosonics_2022}, and declarative sonification grammars~\cite{kim_erie_2024}. Systematic reviews of sonification in assistive systems~\cite{lapusteanu_review_2024} and efforts to reach broader audiences through visualization~\cite{lee_reaching_2020} provide comprehensive overviews of the field's trajectory.

These works collectively demonstrate the potential of tactile, auditory, and multi-modal approaches but remain largely constrained to 2D or static contexts. Three lessons from this body of work directly informed our design approach. First, \textit{multi-modal redundancy}: the principle that information should be encoded through multiple sensory channels simultaneously, guided our integration of sonification, spatial audio, and textual feedback to ensure robust comprehension regardless of individual perceptual preferences~\cite{hoque_accessible_2023, jiang_designing_2024, reinders_when_2025, enge_open_2024, seo_maidr_2024, zong_umwelt_2024}. Second, \textit{pedagogical scaffolding}: as demonstrated in systems that progressively introduce complexity~\cite{seo_maidr_2024, zhang_charta11y_2024, zong_rich_2022}, shaped our inclusion of overview-then-detail (whole to part strategy) features such as multi-perspective auto-play and jump-to-peak navigation. Third, \textit{progressive disclosure}: the idea that systems should allow users to control the granularity of information they encounter~\cite{lundgard_accessible_2021, kim_accessible_2021, sharif_voxlens_2022}, directly motivated our implementation of a switch between point and surface modes which enables users to toggle between aggregated ``squares'' and point-by-point detail and the configurable buffer mechanism. However these 2D methods fail for 3D analytics due to a collapsed topology (squashing three variable correlations into two channel representations), and the lack of occlusion or volumetric traversal. By grounding our 3D visualization prototype in established principles from 2D accessibility research (our multi-modal infrastructure - is inspired by ~\citet{seo_maidr_2024} and ~\citet{seo_maidrai_2024}), we extend their applicability to three-variable correlations and share their core commitment to user autonomy and multi-modal access. Finally, A11yShape~\cite{zhang_a11yshape_2025} is one of the first pieces of literature that allows BLV people to use AI-assist to create 3D models, however it focuses on generative shape modeling through code-based construction (OpenSCAD with GPT-4o) and cross-representation highlighting of discrete geometric components, rather than the systematic, real-time exploration of continuous three-dimensional data visualizations. While A11yShape enables users to \textit{create} and verify geometric forms through AI-generated descriptions and hierarchical abstractions, our work addresses the distinct challenge of \textit{interpreting} pre-existing scientific datasets through features like reference sonification, spatial audio, configurable buffer aggregation, and multi-perspective auto-play that support orientation, comparison, and statistical reasoning within complex data landscapes common to STEM disciplines.

\subsection{Immersive and Embodied Approaches}
\label{subsec:embodiedapproach}
Alongside tactile approaches, a growing body of research explores immersive and embodied interaction as alternative avenues for data access. Extended reality (XR) and embodied interaction research has highlighted the potential of immersive audio and haptic feedback for BLV users~\cite{liu_interactive_2022, liu_research_2020}, with prototypes demonstrating embodied navigation and object labelling in 3D environments. Embodied interaction has been used to support visualization literacy, allowing learners to construct and manipulate visualizations through tangible devices~\cite{johnsonSupportingDataVisualization2023}, embodied allegories for gesture design~\cite{cafaroUsingEmbodiedAllegories2012}, and aesthetics-driven embodied installations~\cite{liTaoistDataVisualization2015}. These studies emphasize the importance of bodily engagement and spatial presence in data interaction. In parallel, immersive environments such as VirtualDesk integrate embodied gestures and virtual workspaces to enhance comfort and analytic performance in 3D settings~\cite{wagnerfilhoVirtualDeskComfortableEfficient2018}, while systematic reviews point to immersive analytics as a rapidly expanding field that seeks to unify spatial, embodied, and multi-sensory techniques~\cite{jamaludinAnsweringWhyWhen2023}.

Although immersive approaches expand the possibilities of non-visual access, they have primarily emphasized proof-of-concept VR or AR experiences for sighted users, rather than evidence-based accessibility for BLV audiences. This gap motivated our discussion of how embodied principles: particularly proprioception, physical movement, and spatial presence; might inform the design of more grounded, accessible interfaces. While our prototype is web-based and keyboard-driven, we deliberately engaged our co-designers in speculative discussions about translating the system into embodied forms (Section~\ref{subsec:exploratory-insights}). These discussions, documented in Session 2, generated critical design insights about advanced haptics, mixed reality synchronization, and alternative physical input devices that both validate embodied approaches as a natural extension of our work and provide concrete pathways for future development beyond screen-based interaction.

\subsection{Gaps in Accessible 3D Data Visualization}
\label{subsec:gapsinaccessible3D}
Despite progress in both tactile and immersive domains, the accessibility of analytic three-dimensional data visualizations remains underexplored. Existing research has developed methods for immersive exploration of 3D scatterplots, addressing challenges like occlusion and density perception through techniques such as Scaptics and Highlight-Planes \cite{prouzeauScapticsHighlightPlanesImmersive2019}, or by investigating movement types and spatial abilities in VR-based scatterplot analysis \cite{simpsonTakeWalkEvaluating}. Collaborative environments such as Shared Surfaces and Spaces highlight how groups interact with 2D and 3D visualizations in immersive contexts \cite{leeSharedSurfacesSpaces2021}, while studies of device-input combinations stress the impact of controllers and display modalities on analytic accuracy \cite{wangUnderstandingDifferencesCombinations2022}. Other work has explored VR systems such as VR-Viz \cite{saifeeVRVizVisualizationSystem2018}, Uncharted Territory’s design heuristics for VR data visualization \cite{UnchartedTerritoryDiving}, and web-based 3D visualization frameworks such as the Digital Library of Mathematical Functions \cite{wangWebbased3DVisualization2005} and dbslice’s infinite canvas approach \cite{pullanVisualizationLargeDatasets}. These systems demonstrate the promise of 3D data interaction across scientific and educational domains but remain largely inaccessible to BLV users. Indeed, even when accessibility is considered, efforts are typically limited to object recognition tasks rather than analytic reasoning with continuous surfaces. 

\begin{table*}[t]
  \centering
  \tiny
  \begin{tabular}{p{3cm} p{4.5cm} p{4.5cm} p{4.5cm}}
    \toprule
    \textbf{Criterion} & \textbf{2D Web-based Encodings} & \textbf{Physical Tactile Displays} & \textbf{Augmented Reality (AR)} \\
    \midrule
    \textbf{Occlusion Handling} &
    Cannot multi-modally represent depth or hidden structures and remain limited to surface projection for 3D surfaces~\cite{farshian_deep-learning-based_2023} &
    Users must physically explore raised lines or pins, so occluded areas require manual navigation~\cite{zhao_tactile_2021} &
    Can simulate 3D depth with audio or visual layering, but hidden or overlapping features may still be difficult to disambiguate for non-visual users~\cite{marquardt_multisensory_2023} \\

    \textbf{Persistence} &
    Encoding via color, glyphs, or text is not appropriate for non-visual access due to high cognitive load~\cite{syvertsen_tangible_2022} &
    Physical relief or pins remain until manually reset, but reconfiguration may be limited to pre-existing states~\cite{je_elevate_2021} &
    AR augmentations persist only while the device is held and tracked; tracking errors or device movement can break spatial consistency~\cite{grimm_vrar_2022} \\

    \textbf{Resolution for Continuous Surfaces} &
    Limited by the granularity of encoding (e.g., color bins, glyph size)~\cite{song_intents_2025} &
    Limited by pin density or embossing precision; highly detailed surfaces are hard to render affordably~\cite{bhatnagar_analysis_2023} &
    High potential: AR can render very fine-grained spatial structure digitally, but non-visual sensing is currently limited~\cite{creed_inclusive_2024} \\

    \textbf{Co-location Requirement} &
    Users can access remotely via personal devices such as laptops and mobile phones~\cite{seo_maidr_2024, zhang_charta11y_2024} &
    User must have the tactile display physically present but do not require co-location with others &
    AR devices (e.g., headset or phone) must be co-located with the physical space or digital model~\cite{geary_design_2022} \\

    \textbf{Cost} &
    Very low: software solutions are inexpensive &
    Often high: refreshable tactile displays or embossers are expensive to build or buy~\cite{mukhiddinov_systematic_2021} &
    High: AR headsets or development of specialized AR for accessibility can be costly~\cite{amer_affordable_2014} \\

    \textbf{Web Integration} &
    Excellent: software and web-native systems are naturally integrated~\cite{seo_maidr_2024} &
    Many tactile displays operate using specialized drivers and often require extensive troubleshooting &
    Moderate to low: AR systems are often standalone applications and not deeply web-integrated~\cite{rafdhi_integrating_2024} \\
    \bottomrule
  \end{tabular}
  \caption{Descriptive comparison of modalities for non-visual or multimodal 3D data access.}
  \label{tab:method-comparison}
  \Description{Comparison table evaluating three modalities for non-visual 3D data access: 2D web-based encodings, physical tactile displays, and augmented reality. Six criteria assessed include occlusion handling, persistence, resolution for continuous surfaces, co-location requirements, cost, and web integration.}
\end{table*}

Although physical tactile systems and AR offer powerful means of perceiving spatial information, they often do not scale well for analytic scenarios (Table ~\ref{tab:method-comparison}). For instance, refreshable tactile displays remain expensive and low-resolution for continuous surfaces (as shown in dynamic tactile marker systems like ~\citet{suzuki_fluxmarker_2017}), while AR solutions for visual impairment frequently depend on co-located headsets or high-end devices, limiting deployment and web integration \cite{coughlan_ar4vi_2017}. Shape-changing interfaces (e.g., pin-array displays) also struggle with cost, actuator density, and inability to dynamically render arbitrary, high-resolution scientific surfaces \cite{leithinger_shape_2015}. Our web-native, multimodal approach avoids these scalability bottlenecks by offering dynamically configurable, low-cost, high-resolution access to 3D surfaces via standard web technologies, enabling widespread and flexible analytic use without requiring any specialized hardware.

Taken together, prior work underscores three trends: (1) accessibility efforts have focused on audiotactile and 2D modalities, (2) immersive and embodied interaction research demonstrates promising but sighted-centered prototypes, and (3) existing 3D visualization systems emphasize technical innovation over inclusive design. To the best of our knowledge, no prior research has applied evidence-based co-design to make three-dimensional data visualizations accessible for BLV users by transferring knowledge from low-fidelity tactile models into high-fidelity, web-native multimodal experiences. This study addresses this gap by integrating tactile and digital probes, empirically evaluating non-visual features such as stereo sonification and volumetric cues, and offering concrete guidance for inclusive 3D visualization workflows.
\section{Co-Design Process and Considerations}
\label{sec:design_process}

Having positioned Experience-Based Co-Design (EBCD) as our methodological foundation in Section~\ref{subsec:accessible2D}, we now detail how we operationalized this framework through concrete design activities. Our adaptation of EBCD's canonical stages (Table~\ref{tab:ebcd-adaptation}) unfolded across two major phases. First, we gathered lived experiences through low-fidelity (\textbf{Phase 2, Thrust 1}) and high-fidelity probes (\textbf{Phase 1 and Phase 2, Thrust 1}). Then, in \textbf{Phase 2 (Thrust 3)}, we held co-design sessions where BLV collaborators prioritized interaction gaps and validated proposed features. Finally, we iteratively implemented solutions grounded in these experiential ``touch-points''. This section documents the structure, composition, and progression of our collaborative inquiry, while Section~\ref{sec:findings} presents the empirical outcomes of these sessions, including the five design goals that emerged from co-designers' narratives and the validated prototype features that these goals motivated.

Our design methodology was rooted in collaborative autoethnography, an approach where researchers engage in a ``relational journey'' to systematically study their own collective experiences~\cite{karalis_noel_collective_2023}. This methodology positioned the design team, particularly its BLV members, as co-participants whose data visualization and research expertise was the very locus of the research. Instead of acting as objective instruments, their role was to engage in dialogic reflexivity, where insights are co-constructed through shared narratives and mutual sense-making. We did not merely design for BLV users; rather, as a mixed-ability team, we systematically documented, analyzed, and responded to our own lived experiences interacting with 3D data visualizations. Our reflections on moments of confusion, discovery, and frustration served as the primary data driving the iterative development of the prototype.

\subsection{Methodological Justification and Design Rationale}
\label{subsec:justification}

We deliberately chose this immersive, autoethnographic~\cite{de_villiers_embodied_2023} EBCD approach~\cite{raynor_experiencebased_2020} over conducting traditional user studies because our research goal was fundamentally generative, not evaluative. A traditional user study framework is optimized for assessing a pre-designed artifact against defined metrics; a process that would have required us to first build a prototype based on our own, likely flawed, assumptions about non-visual interaction. Such a process inevitably centers the researcher's perspective and positions participants as subjects who merely react to a proposed solution. Given that our primary challenge was to translate a deeply embodied, tactile form of 3D data visualization interpretation and discovery into an entirely different digital and auditory modality, a traditional study would have failed to capture the nuanced, iterative process of discovery required. It could tell us if a feature worked, but not why it felt intuitive or disorienting.

In contrast, we adopted an Experience-Based Co-Design (EBCD) approach~\cite{raynor_experiencebased_2020}, which framed our collaborators as co-designers with lived expertise rather than research subjects. As detailed in Section~\ref{subsec:ebcd-mapping}, EBCD treats lived experience as core design material, not merely background context. This methodological orientation shaped our process in several ways. First, our collaborators were engaged from the outset in identifying which challenges mattered and why, with their stories and reflections becoming the starting points for framing design problems grounded in lived realities rather than researcher assumptions. Second, their expertise guided evaluation of our tools not only in terms of usability, but also whether they contributed to autonomy, empowerment, and meaningful access, positioning them as co-judges of value rather than testers of functionality. As established in Section~\ref{subsec:ebcd-mapping}, EBCD distinguishes our work from conventional co-design~\cite{fan_promoting_2025, race_designing_2023} through its epistemological commitment: while co-design often begins from researcher-defined problem spaces and treats participation as a means of generating better solutions, EBCD foregrounds experiential grounding where design work emerges only after lived experiences are surfaced, shared, and co-interpreted, thus reconfiguring whose knowledge counts in defining the problem space. In our project, this meant resisting the urge to ``translate'' existing visualization practices into accessible formats and instead reimagining what visualization could mean when defined through BLV collaborators' perspectives. This shift was essential: it allowed our project to move beyond making visualizations technically accessible, toward exploring how visualization practices themselves might be transformed when co-shaped by BLV experience. Section~\ref{sec:findings} provides concrete examples of how co-designer narratives about orientation confusion and spatial reasoning directly motivated implemented features such as stereo panning, volumetric audio, and reference sonification, demonstrating EBCD's capacity to transform lived experience into grounded design decisions. The following subsections detail the composition of the team whose narratives drove these design trajectories, the specific analytic tasks they addressed, and the structured process through which their expertise was systematically elicited and translated into prototype features.

\subsection{Design Team}
\label{sec:design_team}
Our research was conducted by an interdisciplinary, mixed-ability design team whose collective expertise and lived experiences were foundational to the co-design process. The team integrated seasoned academic scholarship, practical software and hardware engineering, and a rich spectrum of visual abilities, ensuring that our technical explorations were deeply grounded in the nuances of non-visual interaction. 

The core academic group was composed of graduate researchers and faculty who drove the project’s theoretical framing and iterative development. Sanchita, a sighted Graduate Teaching and Research Assistant with a minor in Data Analytics, led the technical implementation and interface development. She leveraged extensive prior experience creating 3D visualizations in MATLAB and a strong understanding of statistical relationships to support the system’s analytical foundations. Aziz, a Graduate Teaching and Research Assistant who is blind, provided continuous, real-time feedback from the perspective of a user with a foundational understanding of data visualization. Aziz has experience reading and creating common plot types: including bar, line, box, scatter plots, histograms, and heatmaps; using visual, tactile, and sonified approaches, and developed his 3D visualization knowledge through participation in this project.
They were joined by Venkatesh, an Assistant Professor of Information with seven years of research experience in data visualization and programming tool accessibility, who contributed intermediate-to-advanced expertise shaped by his doctoral training and ongoing research in the field. Sile, a Professor of Information with 25 years of teaching experience, helped shape the cognitive aspects of this project. She regularly analyzes data in her own research and has contributed to preparing instructional materials through initiatives such as the Summer Institute. JooYoung, an Assistant Professor of Information Sciences, brought over eight years of expert knowledge in creating accessible and multimodal data visualization tools. His work includes the development of R packages for accessible visualization and deep expertise in statistical concepts and research methods. This faculty trio guided two graduate researchers who were central to the prototype’s creation. 

To bridge academic theory with real-world application, the team was expanded to include Ken, a Senior Software Engineer and R\&D specialist in assistive technology. Ken contributed extensive expertise spanning both software and hardware, including the design of tactile graphics displays, paper tactile graphics, braille-to-digital pipelines, audio–tactile multimodal visualization systems, and data exploration workflows using tools such as pandas, Seaborn, and Matplotlib. He has also developed 3D tactile displays and collaborated with 3D printing workflows to represent STEM data, including exploratory work with haptic and tactile pin-array systems. He provided a critical perspective on the practical needs of blind and low-vision learners beyond the laboratory.
Crucially, the team’s strength was rooted in its diversity of lived experience. Five of the six members (JooYoung Seo, Sile O’Modhrain, Venkatesh Potluri, Aziz Zeidieh, and Ken Perry) are blind or have low vision, with visual acuities represented in Table~\ref{tab:visual-acuity}. This rich tapestry of perspectives was the engine of our collaborative autoethnography, positioning BLV researchers as the primary research actors and ensuring that the design process was driven not by assumptions, but by a nuanced, collective understanding of non-visual data interaction.

\begin{table*}[ht]
\centering
\scriptsize
\caption{Team members’ visual acuity, onset of impairment, age, and roles.}
\label{tab:visual-acuity}
\begin{tabular}{p{0.8cm} p{3.5cm} p{1.5cm} p{0.5cm} p{4cm}}
\hline
\textbf{Name} & \textbf{Visual Acuity} & \textbf{Age of Onset} & \textbf{Age} & \textbf{Occupation} \\
\hline
Sanchita & Not BLV; corrected to 20/20 with lenses & Not applicable & 24 & Graduate Teaching and Research Assistant \\
Aziz & No vision in left eye; 20/2000 in right eye & Congenital & 28 & Graduate Teaching and Research Assistant \\
Venkatesh & Totally Blind & Congenital & 32 & Assistant Professor of Information \\
Sile & Some light perception & Congenital & 59 & Professor of Information \\
Ken & Totally Blind & 20 & 55 & Senior Software Engineer and R\&D Specialist \\
JooYoung & No light or shape perception in right eye, light and shape perception in left eye & 11 & 35 & Assistant Professor of Information Sciences \\
\hline
\end{tabular}
\end{table*}

\subsection{Design Process}
\label{sec:design_process}
Our inquiry was methodologically structured through a sequence of interconnected stages, inspired by the EBCD Framework~\cite{raynor_experiencebased_2020}. We adapted its healthcare-centric model by translating the ``patient pathway'' into the user's journey of exploring a 3D visualization. \textbf{Phase 1}, (May to June 2025), initiated the project with a foundational ideation design cycle involving BLV and non-BLV collaborators (JooYoung, Aziz, and Sanchita), which produced an initial high-fidelity digital probe. The research then progressed to \textbf{Phase 2} (July to August 2025), organized into three distinct thrusts to systematically refine this probe. \textbf{Thrust 1} (first week of July 2025) involved the fabrication of a physical, low-fidelity probe to serve as an embodied reference. In \textbf{Thrust 2} (July 2025), insights from the end of \textbf{Phase 1} were integrated into the high-fidelity digital probe. Finally, \textbf{Thrust 3} which expanded the collaborative team to include BLV co-designers (Sile, Venkatesh, Ken) and comprised two iterative co-design sessions lasting 90 minutes each: \textbf{Session 1} (12 Aug 2025) centered on a comparative analysis of the experience provided by two probes to identify interaction gaps (mostly situated around knowledge transfer from physical to digital prototype), while \textbf{Session 2} (22 Aug 2025) was dedicated to testing and validating new features that emerged from this analysis, culminating in future directions that could provide the platform to observe embodied learning and interactions of users.

\begin{figure}
    \centering
    \includegraphics[width=\linewidth]{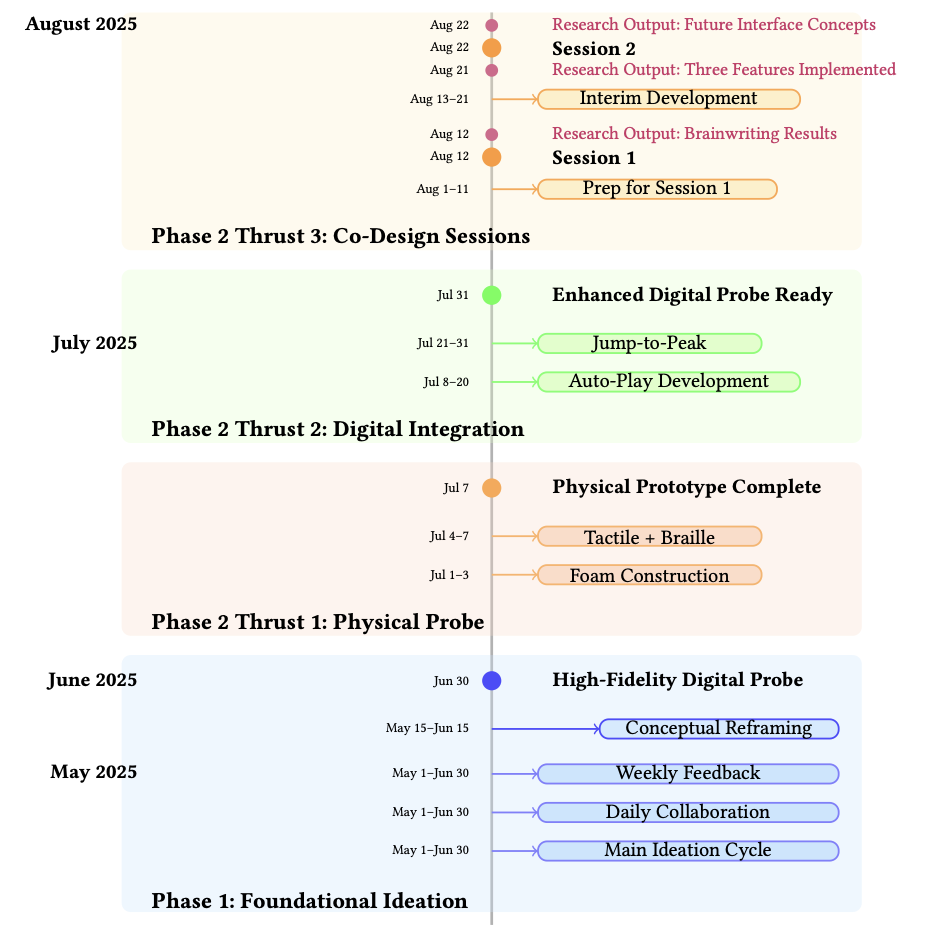}
    \caption{Experience-Based Co-Design: Research Timeline}
    \Description{Gantt chart spanning May-August 2025 showing five research phases. Phase 1 (May-June): Foundational ideation with JooYoung, Aziz, and Sanchita producing initial digital probe. Phase 2 Thrust 1 (July 1-7): Physical tactile probe fabrication. Thrust 2 (July 8-31): Digital integration implementing auto-play and jump-to-peak features. Thrust 3 (August 1-22): Two co-design sessions with all BLV researchers; Session 1 identifies four features, Session 2 validates three implementations and explores future embodied interfaces. Milestones marked with diamonds track prototype evolution and design insights.}
    \label{fig:research-timeline}
\end{figure}

\subsubsection{\textbf{Phase 1: Foundational Ideation and High-Fidelity Probe Development}}
\label{subsubsec:phase1}
\textbf{Phase 1} was grounded in a mixed-ability co-design methodology, drawing upon the interdependence framework~\cite{bennett_interdependence_2018} to structure a sustained, two month long collaboration between two blind researchers (JooYoung, Aziz) and one non-BLV researcher (Sanchita) for creating an initial version of our high-fidelity design probe. In this initial phase, JooYoung and Aziz provided crucial design insight rooted in their lived experience, while Sanchita led the implementation and interface development. The process was highly iterative, characterized by daily collaboration with Aziz and weekly sessions with JooYoung, where each meeting involved direct evaluation of the evolving probe, the proposal of new features, and the identification of interaction pain points. This continuous feedback loop allowed Sanchita to implement refinements in real-time. A central activity of this phase was the conceptual reframing of conventional visual affordances, such as grid overlays and highlighting, into non-visual strategies like wireframe traversal, axis-specific navigation, and the synchronized alignment of auditory, visual, and textual feedback. This foundational work culminated in the production of the initial high-fidelity digital design probe, which served as the empirically grounded starting point for the expanded research activities in \textbf{Phase 2} - \textbf{Thrust 2} and \textbf{Thrust 3}.

\subsubsection{\textbf{Phase 2: Evidence-Based (Re)Design and Knowledge Transfer}}
\label{subsubsec:phase2}
The second phase of the research was designed to rigorously test, critique, and reconstruct the initial probe. The overarching goal of \textbf{Phase 2} was to convert the high-fidelity probe into a functional prototype capable of facilitating effective knowledge transfer from a physical, embodied medium to a digital, multi-modal environment. This work was organized into three distinct but interconnected thrusts, with Sanchita, Aziz, and JooYoung leading the first two.

\textbf{Thrust 1} of \textbf{Phase 2} was the creation of a physical, low-fidelity probe. This probe functioned not merely as a demonstrative tool but as a crucial epistemological anchor; it established a ``ground truth'' of embodied understanding, providing a shared sensory baseline against which the digital experience could be systematically evaluated.
\textbf{Thrust 2} of \textbf{Phase 2} focused on implementing initial insights from our foundational work. Prior to engaging the broader group, Sanchita integrated the key design recommendations from JooYoung, which had been identified in \textbf{Phase 1}, into the digital probe. This act of translation ensured methodological continuity and established a starting point for the expanded sessions that was already grounded in expert user knowledge.
\textbf{Thrust 3}, the most extensive component of \textbf{Phase 2}, was the execution of two iterative co-design sessions.
\begin{enumerate}
    \item \textbf{Session 1:} This session was structured as a direct comparative analysis. JooYoung, Sile, Venkatesh, and Ken were invited to first explore the physical probe from \textbf{Thrust 1}, using manual and tactile inspection to construct a direct understanding of the data's form by feeling its gradients, slopes, and peaks. Subsequently, they transitioned to the high-fidelity digital probe from \textbf{Thrust 2}, using keyboard-only navigation and sonification to explore the same data structure. Session 1 concluded with a group discussion and a silent brainwriting exercise whose prompts were designed by Sanchita (readable at Appendix Section~\ref{subsec-app:session1-bwprompt}). This exercise produced a clear consensus on four major features required to bridge this experiential gap which we will talk about in Section \ref{sec:findings}.
    \item \textbf{Interim Development:} During the interim period between Session 1 and 2, Aziz and Sanchita (posing themselves as co-developers - with Sanchita leading implementation and Aziz providing iterative feedback) implemented three of the four prioritized features. 
    \item \textbf{Session 2:} This session was bifurcated into two distinct parts. The first part was dedicated to the systematic testing and validation of three of the four prioritized features implemented after Session 1 (Section~\ref{subsubsec:findings_phase2-session2-part1}).
    The second part of the session shifted to future-oriented ideation, where collaborators engaged in a brainwriting exercise to brainstorm how the embodied knowledge from the physical prototype could be transferred into more immersive digital environments (Section~\ref{subsubsec:findings_phase2-session2-part2}). Session 2 concluded with a collaborative brainwriting and discussion exercise to generate speculative ideas that could generate an embodied 3D visualization learning experience. The prompts were designed by Sanchita (readable at Appendix Section~\ref{subsec-app:session2-bwprompt}).
\end{enumerate}

\subsubsection{Analytic Tasks and Co-Design Prompts} \label{subsubsec:analytic_tasks}
In these Sessions, we explicitly targeted core analytic tasks essential for nonvisual 3D data exploration, tasks that are foundational to scientific reasoning across STEM disciplines. Our co-designers were prompted to engage with the following challenges during both probe exploration sessions and the brainwriting exercises:
\begin{itemize}
    \item \textbf{3D Orientation:} Establishing and maintaining awareness of position and direction within the three-dimensional coordinate space; prompts included ``How do you know where you are in the plot?'' and ``What cues help you understand the axes?''
    \item \textbf{Landmark and Peak Finding:} Identifying local maxima, minima, and other salient features; prompts included ``How would you locate the highest point?'' and ``Can you find where the data changes most dramatically?''
    \item \textbf{Comparing Local Maxima versus Global Trends:} Distinguishing between isolated peaks and broader patterns; prompts included ``How do you differentiate a single peak from an overall rising trend?'' and ``What would help you compare multiple high points?''
    \item \textbf{Gradient Tracing:} Following the direction and steepness of slopes across the surface; prompts included ``Can you follow the path of steepest ascent?'' and ``How would you trace a ridge or valley?''
    \item \textbf{Identifying Occluded or Partially Hidden Features:} Detecting regions obscured by overlapping data or structural complexity; prompts included ``How do you explore areas that might be hidden from one perspective?'' and ``What strategies help you uncover features that aren't immediately apparent such as those with similar Y values?''
\end{itemize}
These tasks were not presented as formal test scenarios but rather as conversational prompts woven into our co-design sessions, encouraging co-designers to articulate their exploration strategies, confusions, and moments of insight. The brainwriting exercises at the end of Session 1 and Session 2 (Appendix Sections~\ref{subsec-app:session1-bwprompt} and~\ref{subsec-app:session2-bwprompt}) further formalized these prompts, asking co-designers to reflect on which features supported or hindered each task. The complete set of task-based questions used throughout both sessions is documented in Appendix Sections~\ref{subsec-app:session1-tasks} and~\ref{subsec-app:session2-tasks}, connecting the five analytic tasks directly to the specific prompts and evaluation criteria employed during probe exploration and feature validation.
The brainwriting exercise in Session 1 produced a clear consensus on four major features, but the path from co-designer narratives to these implemented features merits closer examination. Two such examples illustrate this translation process. During probe exploration, JooYoung articulated a fundamental analytic limitation: the difficulty of comparing different regions within a single plot. Where the physical probe allowed simultaneous tactile comparison using multiple fingers, the digital interface forced sequential, memory-dependent exploration. Their narrative directly motivated the configurable buffer feature, implemented during the interim development period, which enables users to save a region's sonification and compare it against their current focus point. Similarly, when co-designers reported spatial disorientation, Ken and Sile proposed adding reference sonifications. These narratives coalesced into the fixed origin sonification with replay mechanism, validated in Session 2. These trajectories exemplify how EBCD centered co-designers' experiential knowledge to produce features addressing real, lived challenges rather than theoretically anticipated needs.

\urlstyle{tt}
\section{Design Probes}
\label{sec:design-probes}

\subsection{Low-Fidelity Design Probe}
\label{subsec:lowfi_design-probe}
\begin{figure}[h!]
    \centering
    \includegraphics[width=0.7\linewidth]{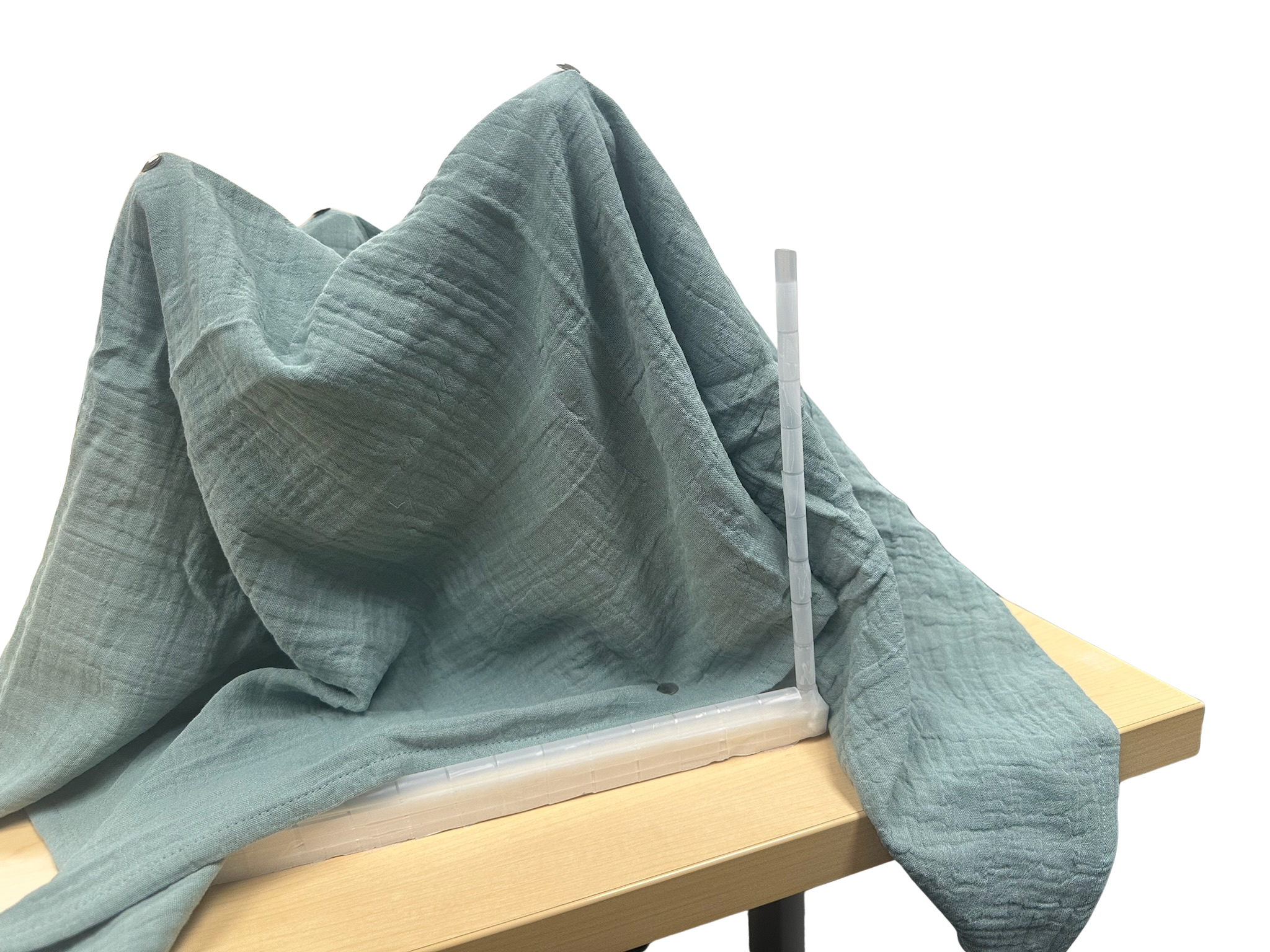}
    \caption{Low-Fidelity Prototype}
    \Description{Tactile 3D surface plot on polystyrene foam base (11.7 x 16.5 inches) covered with teal muslin cloth draped to form peaks and valleys. Plastic tubing forms L-shaped axes with braille labels: vertical Y-axis with tape-marked ticks, horizontal Z-axis and X-axis along edges. Thin tubes of varying heights topped with aluminum foil balls represent data points beneath the cloth. The fabric transforms discrete points into a continuous tactile surface for exploring topography through touch.}
    \label{fig:low-fi}
\end{figure}

The low-fidelity probe (created in \textbf{Phase 2, Thrust 1} and employed in \textbf{Phase 2, Thrust 3, Session 1}) was constructed on an 11.7 x 16.5 inch, 3/16-inch thick polystyrene foam sheet, which formed a sturdy base and provided the co-designers with a clear orientation for the plot. We affixed thicker plastic tubes (from 19.69-inch balloon sticks - marked with tape to represent the ticks) along two perpendicular edges to represent the Z-axis (longer edge) and X-axis (shorter edge). A small peg holder was attached at the origin for the co-designers to insert a separate tube for the Y-axis, perpendicular to the base. To represent the raw data (5 points), an array of peg holders was attached across the foam board; into these, we inserted thinner plastic tubes cut to varying lengths corresponding to the height of each data point. Small balls of heavy-duty aluminum foil were placed on the tip of each tube to create a distinct tactile representation of the points. All elements, including the axes, were labeled using a 6-dot braille label maker for full accessibility. Finally, a 38 x 40 inch muslin cotton cloth was draped over the points to simulate the continuous mesh of the surface plot. Developed by Sanchita, this tactile artifact translated the abstract, visual topography of a 3D surface plot into a tangible landscape of peaks, valleys, and slopes (viewable in Figure~\ref{fig:low-fi}). 

\subsection{High-Fidelity Design Probe}
\label{subsec:hifi_design-probe}
Our high-fidelity design probe (created in \textbf{Phase 2, Thrust 2} and employed in \textbf{Phase 2 Thrust 3, Session 1 and 2}) is a browser-based, multi-modal platform that evolved directly from our auto-ethnographic inquiry. To facilitate this, we made a deliberate methodological choice to build a modular system architecture. This architectural separation was the technical manifestation of our EBCD methodology; it was chosen specifically because it allowed us to be immediately responsive to collaborator feedback. By isolating concerns, we could rapidly prototype and integrate new, complex features—often within a single layer—without destabilizing the entire application. This enabled us to translate the insights from our co-design sessions directly into functional system logic.

The system is composed of five strictly hierarchical layers that communicate through an event-driven architecture using centralized event constants (defined in \url{EventConstants.js}):
\begin{itemize}
    \item \textbf{Data Layer} (\url{PlotData.js}): Manages data ingestion, validation, normalization, and statistical analysis.
    \item \textbf{Engine Layer} (\url{VisualizationEngine.js}): Handles WebGL rendering for points, surfaces, and contours using custom GLSL shaders.
    \item \textbf{Accessibility Layer} (\url{NavigationController.js}): Contains all non-visual interaction logic coordinating eleven controllers for sonification, text feedback, navigation, autoplay, jump-to-peak, drag-select, and review modes.
    \item \textbf{UI Layer} (\url{UIController.js}): Manages visual interface components including menus, color schemes, axes rendering, and AI chat.
    \item \textbf{Application Layer} (\url{app.js}): Serves as global coordinator, initializing dependencies and orchestrating cross-layer communication.
\end{itemize}

Layer communication follows strict architectural constraints: components access only their own layer and lower layers through dependency injection, while cross-layer events propagate through a custom event bus. Representative event payloads include \url{display-mode-changed} (notifying mode switches), \url{drag-select-selection-confirmed} (communicating buffer boundaries), and \url{autoplay-state-changed} (coordinating playback status). This architecture enabled us to implement complex features like buffer aggregation and jump-to-peak functionality within isolated controllers while maintaining system stability across iterative (re)designs.

\paragraph{Sonification Engine and Web Audio Implementation}\label{subsubsec:sonification-implementation}
The sonification engine (\url{SonificationController.js}) is implemented as a custom Web Audio API synthesis pipeline operating at the browser's native sample rate (typically 48kHz). Audio synthesis occurs entirely client-side without external libraries, using the \url{AudioContext} interface to create real-time parameter-mapped oscillators. The core signal path begins with oscillator generation (\url{createOscillator()}), routes through envelope shaping (\url{createGain()}), applies spatial positioning via binaural panning (\url{createStereoPanner()}), and adds depth perception through convolution reverb (\url{createConvolver()}).
Timbre encoding employs three oscillator waveforms: sine for smooth low-frequency regions, triangle for mid-range transitions, and square for high-frequency accents; selected dynamically based on normalized X-axis position. The binaural setup maps X-coordinates to stereo pan values (-1.0 for leftmost to +1.0 for rightmost), creating horizontal spatial positioning that requires headphone playback for accurate localization. Depth perception (Z-axis) is rendered through algorithmic reverb: we generate a 2.2-second impulse response buffer with exponential decay (power 3.5) and apply variable wet/dry mixing (20-95\% wet) scaled by normalized depth values. Pre-delay ranges from 10ms (foreground) to 90ms (background), with lowpass filtering (6500Hz to 2000Hz cutoff) simulating distance-dependent damping. Round-trip latency from keystroke to audio output measures approximately 15-30ms on modern browsers, below the threshold for perceived interaction delay.

\paragraph{Visual Highlighting System and WebGL Implementation}\label{subsubsec:visual-highlighting-implementation}
The visual highlighting system (\url{HighlightController.js}) provides high-contrast visual feedback for navigation and selection through direct manipulation of WebGL rendering buffers. The controller operates by maintaining state for three distinct highlighting contexts: single-point navigation (magenta highlight with 4x size multiplier), multi-selection regions (white highlighting for drag-selected areas), and cursor positioning (dynamic yellow/white coloring based on selection state). The core highlighting mechanism integrates with the Engine Layer's \url{VisualizationEngine.js} through the \url{createBuffers()} pipeline: when a point or wireframe rectangle is marked for highlighting via \url{setHighlightedPoint(index)} or \url{setHighlightedWireframeRectangle(index)}, the controller triggers complete buffer regeneration, during which the rendering engine queries \url{isPointHighlighted(index)} for each data point to determine if enhanced visual properties should be applied.
For point-mode visualization, highlighted vertices receive modified RGBA color arrays (retrieved via \url{getHighlightColor()}) and scaled point sizes (computed through \url{getHighlightSizeMultiplier()}) before being uploaded to GPU vertex buffers using WebGL's \url{gl.bufferData()} with \url{STATIC\_DRAW} usage pattern. In surface mode, the highlighting system employs separate WebGL buffer objects (\url{highlightPosition}, \url{highlightColor}, \url{highlightIndices}) to isolate highlighted wireframe rectangles from the base mesh geometry. During rendering, these buffers are drawn with aggressive depth handling; disabling depth testing (\url{gl.disable(gl.DEPTH\_TEST)}), applying polygon offset (\url{gl.polygonOffset(-1.0, -1.0)}), and forcing opaque rendering through the simpler line shader pipeline to ensure highlighted elements render atop the base visualization regardless of y-ordering. This architecture maintains strict separation between highlighting logic (Accessibility Layer) and rendering implementation (Engine Layer), enabling the system to support simultaneous multi-selection highlighting, cursor feedback, and navigation focus without buffer conflicts.

\subsubsection{\textbf{Initial System: Before Session 1}}
\label{subsubsec:before-session1}
The initial version of the probe was developed during and after the foundational work in \textbf{Phase 1} (read more in Section~\ref{subsec:findings_phase1}), with features aimed at enhancing user autonomy. To address the tedium of manual traversal, we implemented two key exploration strategies within the Accessibility Layer:
\begin{itemize}
    \item Multi-Perspective Auto-Play (\emph{AutoPlayController.js}): This controller implements our ``whole-to-part'' navigation strategy by sequencing navigation commands and triggering corresponding sonification events, allowing users to receive auditory ``screenshots'' of the data without exhaustive manual traversal.
    \item Jump-to-Peak Function (\emph{JumpController.js}): This feature statistically analyzes the data (view paragraph below for a detailed explanation) to identify the most significant peaks and troughs, enabling rapid navigation between a surface plot's most salient features with a simple \emph{J key} press.
\end{itemize}
In addition to the core visualization features, Sanchita implemented a multi-threaded AI chat assistant that leverages Gemini 2.5 Pro to support interactive data exploration. This feature integrates directly with the Visualization Engine, enabling users to query the underlying dataset or trends visible in the 3D plot. However, this feature was not extensively tested during \textbf{Session 1} or \textbf{Session 2}, as our primary research focus remained on independent, non-visual data exploration and the transfer of knowledge between low-fidelity and high-fidelity design probes. A detailed description of the AI chat assistant implementation is provided in Appendix~\ref{subsec-app:ai-chat}.

\paragraph{Jump-to-Peak Statistical Algorithm}\label{subsubsec:jump-algorithm}
The Jump-to-Peak feature (\url{JumpController.js}) identifies salient features through statistical outlier detection operating on wireframe rectangles. The algorithm: (1) extracts \url{avgY} values from all rectangles; (2) sorts by \url{avgY} to select top 20 positive and bottom 20 negative candidates; (3) removes spatial duplicates by \url{(x,z,avgY)} key; (4) calculates dataset statistics (\url{mean}, \url{min}, \url{max}, \url{range}); (5) computes 20\% threshold $= (\texttt{max} - \texttt{min}) \times 0.2$; (6) tests proximity: if negatives are within threshold but positives are not, selects top 10 positive peaks only; if positives are within threshold but negatives are not, selects bottom 10 negative peaks only; otherwise selects top 10 of each; (7) sorts final peaks with positives descending and negatives ascending; (8) returns array of peak objects with sign and coordinates.
When activated by pressing J in surface mode, the controller saves the current navigation position, detects peaks using this algorithm, and sequentially jumps through the identified features. Each jump updates the grid position to align with the peak rectangle's center coordinates, highlights the rectangle visually, and plays a distinctive frequency-swept tone (400Hz base for negative peaks, 800Hz for positive peaks). Pressing J again cycles to the next peak, while Escape exits the mode and restores the saved navigation position.

\subsubsection{\textbf{System (Re)Design: After Session 1}}
\label{subsubsec:after-session1}
Based on Sile, Venkatesh, Ken and JooYoung's interaction with both low-fidelity and high-fidelity design probes, we postulated some design features to be implemented (read more in-depth in Section~\ref{subsec:findings_phase2-session1}).
To address the critical need for stable orientation, we enhanced the \emph{SonificationController.js} within the \emph{Accessibility Layer} to provide a fixed reference sound for the origin (0,0,0) by pressing the \emph{0 key} and a replay sonification mechanism of current point by pressing \emph{. key}. 
The reference sound acts as a constant, while the replay function, bound to the \emph{. key}, allowing a user to re-listen to their current location's sonification without moving. This same controller is also responsible for encoding position and depth cues through stereo panning and volumetric spatial audio. 
As a user navigates along the X-axis, the sound pans between the left and right stereo channels; movement along the Z-axis is conveyed through changes in volume, creating an intuitive auditory analog for proximity and distance. 
Finally, to balance analytic detail with cognitive load; another key finding from \textbf{Session 1} (read in depth in Section~\ref{subsec:findings_phase2-session1}), we developed \url{DragSelectController.js} to implement configurable buffer playback as detailed below. The implementation spanned multiple layers: the Accessibility Layer managed selection logic and queried the Data Layer to retrieve buffered points or rectangles, calculated arithmetic means for aggregated mode, and organized sequential mode playback. The \url{HighlightController.js} (Accessibility Layer) coordinated with \url{UIController.js} (UI Layer) to render visual feedback through \url{VisualizationEngine.js} (Engine Layer). This empowers users to fluidly shift between high-level summaries and detailed inspections of specific areas, streamlining the discovery process. Users could upload their own data as well, and all described multi-modal features are extended to the rendered plot.

\paragraph{Buffer Aggregation and Playback Modes}\label{subsubsec:buffer-aggregation}
Buffer selection (\url{DragSelectController.js}) implements a three-step rectangular region selection: users press D to enter selection mode, then press Enter to begin anchoring the selection start point at their current navigation position. They navigate using arrow keys to define the opposite corner (visual highlighting also follows the expanding selection area), then press Enter again to confirm and store the region. The controller queries the Data Layer for all points or wireframe rectangles with coordinates falling within the defined bounds, storing them in \url{currentSelection} as a buffer array that persists until a new selection is made.
Aggregation applies arithmetic mean calculation: for each selected item, we normalize coordinates (handling both point objects with x/y/z properties and rectangle objects with centerX/avgY/centerZ properties), sum all valid Y-values, and divide by count to produce \url{averageYValue}. The aggregated playback mode (playable though the G-key) sonifies this single mean value for 1.0 second duration. In sequential (non-aggregated) mode, the buffer is organized by Z-index rows (front-to-back) with left-to-right X-coordinate sorting within each row. Each item plays for 0.3 seconds with 125ms inter-item silence, creating a systematic auditory sweep. Users toggle between modes by pressing G: the first press activates sequential playback, the second switches to aggregated, and the third returns to sequential.
\section{Findings}
\label{sec:findings}
Our findings are drawn directly from the auto-ethnographic data collected during the end of \textbf{Phase 1} and the two co-design sessions in \textbf{Phase 2, Thrust 3}. 

\subsection{\textbf{End of Phase 1: Allowing User Autonomy over Data Exploration}}
\label{subsec:findings_phase1}
Conversations and iterative testing with JooYoung revealed that empowering users with autonomy was the most critical factor for successful data exploration, consistent with prior findings in accessible visualization that emphasize user-controlled interaction and progressive disclosure~\cite{lundgard_accessible_2021, kim_accessible_2021}. They expressed that a purely manual, point-by-point traversal of the data was not only tedious but also made it difficult to build a cohesive mental model of the 3D plot, echoing challenges reported in sequential non-visual exploration of complex data~\cite{sharif_understanding_2021}. Their feedback consistently highlighted the need for tools that would allow them to control the scope and focus of their exploration, balancing high-level overviews with the ability to investigate specific features efficiently. This led to the development of our first two design goals aimed at granting users greater control over their interaction with the data.
\begin{enumerate}
    \item [DG1.]\textbf{\textit{ The prototype should provide multi-perspective auto-play to deliver auditory ``snapshots'' of the dataset:}} Co-designers emphasized the need for rapid overviews of complex plots; auto-play from multiple perspectives was valued as a way to capture structural patterns across all three axes without exhaustive manual traversal, aligning with overview-first strategies in accessible visualization~\cite{seo_maidr_2024, zong_rich_2022}.
    \item [DG2.]\textbf{\textit{ The prototype should include a jump-to-peak function to quickly locate salient features: }} JooYoung highlighted the value of a dedicated command for automatically cycling between local maximum peak and trough values, enabling efficient interrogation of surface plot trends without exhaustive searching, similar to landmark-based navigation approaches~\cite{zong_umwelt_2024, sharif_voxlens_2022}.
\end{enumerate}
  
\subsection{\textbf{Phase 2, Session 1: Surfacing the Experiential Gap}}
\label{subsec:findings_phase2-session1}
\textbf{Session 1} directly addressed \textbf{RQ1} by revealing a fundamental disconnect between the tactile prototype's physical understanding and the digital version's abstract experience, a gap previously observed when translating tactile or embodied representations into screen-based systems~\cite{vandemoerePhysicalVisualizationInformation2010, hermanTouchingGroundEvaluating2025}. Co-designers immediately surfaced multiple orientation and comparison challenges that mapped directly onto five essential analytic tasks: 3D orientation, gradient tracing, comparing local maxima versus global trends, landmark and peak finding, and identifying occluded features. Table~\ref{tab:session1-eval} systematizes the prompts we used to elicit these experiences, the analytic tasks they addressed, the observed issues that emerged from co-designer narratives, and the design changes we subsequently implemented to support each task.

\begin{table*}[t]
  \centering
  \scriptsize
  \begin{tabular}{p{3.5cm} p{2cm} p{4cm} p{4cm}}
    \toprule
    \textbf{Prompt} & \textbf{Analytic Task} & \textbf{Observed Issue} & \textbf{Design Change Implemented} \\
    \midrule
    ``How do you know where you are in the plot? What cues help you understand the axes?'' & 3D Orientation & Axis convention misalignment: co-designers expected Y-axis on board, Z-axis pointing out; digital used Y-vertical, Z-depth. Ken: ``I would expect the y-axis to be on the board and the z-axis pointing out.'' Venkatesh: ``X-axis would be along the horizontal edge of the board, Y-axis would be along the vertical edge... and Z-axis would be the one popping.'' & Fixed reference sonification at origin (0,0,0) via 0-key; replay mechanism via .-key to re-hear current position (\textit{DG3}). \\
    \midrule
    ``Can you follow the path of steepest ascent? How would you trace a ridge or valley?'' & Gradient Tracing & Lack of spatial encoding for horizontal and depth dimensions. Venkatesh: ``it would be nice to have it pan from left to right, and decrease the volume as it's going forward.'' & Stereo panning for X-axis (left-right), volumetric audio (volume + reverb) for Z-axis depth (\textit{DG4}). \\
    \midrule
    ``What would help you compare multiple high points?'' & Comparing Local Maxima vs. Global Trends & Difficulty maintaining mental representation of previous positions during sequential exploration; physical probe allowed simultaneous tactile comparison. & Configurable buffer: drag-select to save region, toggle between aggregated (mean) and sequential (point-by-point) playback via G-key (\textit{DG5}). \\
    \midrule
    ``How would you explore areas that might be hidden from one perspective?'' & Identifying Occluded Features & Inter-plot comparison not supported; need to differentiate datasets by categorical variable. Ken: ``What if I wanted to compare two sets of data... Is there any way to overlay them and switch between the two surfaces?'' Sile: ``if you gave the sound for Georgia a square wave instead of a sine wave.'' & Identified for future implementation: dual-dataset overlay with distinct timbres and keybinding toggle. \\
    \bottomrule
  \end{tabular}
  \caption{Session 1 evaluation: prompts, analytic tasks, observed issues, and implemented design changes.}
  \label{tab:session1-eval}
  \Description{Table documenting Session 1 findings across four analytic tasks. Each row maps evaluation prompts to observed issues and design solutions: 3D orientation issues led to reference sonification; gradient tracing difficulties prompted stereo panning and volumetric audio; comparison challenges motivated configurable buffers; occlusion exploration needs identified future dual-dataset overlay requirements. Includes direct quotes from co-designers Venkatesh, Ken, and Sile supporting each design decision.}
\end{table*}

A primary point of friction was the orientation of the coordinate system, which we anticipated from JooYoung's experience in \textbf{Phase 1}. However, we wanted to gather insight from Sile, Venkatesh and Ken before we implemented features, so that we could learn from their expertise and follow the EBCD framework guidelines with all our collaborators~\cite{morley_evidence-informed_2024}. Notably, this issue surfaced even before the co-designers interacted with our digital prototype; participants expressed expectations about axis orientation while experiencing the physical tactile probe, consistent with prior work on embodied spatial expectations~\cite{johnsonSupportingDataVisualization2023}. For example, Ken remarked, ``Yeah, I find that kind of odd, too. I would expect the y-axis to be on the board and the z-axis pointing out.'' Venkatesh shared a similar mental model: ``...what I was expecting is X-axis would be along the horizontal edge of the board, Y-axis would be along the vertical edge... and Z-axis would be the one popping.''
These comments referred specifically to their expectations for the physical device, but the same mental model carried over when they later interacted with the digital prototype, whose coordinate system followed standard visualization conventions (vertical Y, horizontal X, depth Z). This mismatch helped us identify the need to explicitly align coordinate orientations across both physical and digital components.

The (re)design process was central to our methodology, representing a direct, evidence-based response to the core challenges identified during our co-design sessions. The impetus for the (re)design was the experiential gap that collaborators surfaced when comparing the low-fidelity tactile probe with the initial high-fidelity digital version, reflecting similar tensions observed in prior tactile-to-digital translation efforts~\cite{syvertsen_tangible_2022}. 
While the tactile model afforded a rich, tactile understanding, the digital probe felt ``abstract'' and ``disoriented'', necessitating a fundamental reconstruction. This redesign involved implementing empirically grounded features to bridge this gap, including a fixed reference sonification for orientation, stereo and volumetric audio to encode position~\cite{siu_virtual_2020, hoque_accessible_2023}, and a configurable buffer playback to balance analytic detail with cognitive load. Conceptually, this (re)design process was our primary method for beginning to translate tactile knowledge into a web-based digital form. It signifies a shift from creating a static artifact to engaging in a dynamic, collaborative refinement aimed at enhancing users' analytical agency and empowering them to independently interpret complex 3D data.

Venkatesh proposed using spatial audio, suggesting, ``it would be nice to have it pan from left to right, and decrease the volume as it's going forward.'' This idea of mapping spatial dimensions to audio properties was a recurring theme and formed the basis for our efforts to incorporate stereo and volumetric spatial audio to encode position and depth cues.
The concept of a stable reference point also emerged as a critical need. Ken suggested, ``You could even mark a reference and then walk away from it on the surface''. Sile expanded on this, proposing a dedicated key binding: ``You could have a reference just available on a key, where you would just ping it''. This exchange was the genesis of our goal to provide a reference point and replay mechanism to restore orientation during exploration.
A significant limitation of the initial digital probe was the difficulty of comparing different points or regions within a single plot. This led to the idea of a ``custom buffer''. As JooYoung proposed, ``...it might be instrumental, if we have a, like, custom buffer, so that, say, for example, we can save one specific segment in specific buffers... And then you can hear that buffer... and you can also replay your currently focused segment''. This concept of selecting and saving a region for later comparison formed the basis for our goal to implement configurable buffer playback to balance analytic detail and cognitive load.

Beyond intra-plot comparison, the discussion surfaced the need for inter-plot comparison. Ken raised a critical use case: ``What if I wanted to… Compare two sets of data, same surface types... Is there any way to overlay them and… Switch between the two surfaces?'' This question identified a fourth key feature: the ability to compare two datasets differentiated by a categorical variable (e.g., precipitation in Texas vs. Georgia). The group brainstormed solutions, with Sile suggesting the use of distinct timbres to differentiate the two plots: ``if you gave the sound for Georgia a square wave instead of a sine wave''. This idea of layering two datasets in the same coordinate space and allowing a user to switch between them using a keybinding, with distinct auditory feedback for each, was identified as a crucial direction for future development, even though it was not implemented before \textbf{Session 2}. Thus, our final three design goals were formulated:
\begin{itemize}
  \item [DG3.]\textbf{\textit{ The prototype should provide a reference point and replay mechanism to restore orientation during exploration:}} Co-designers identified the fixed origin sonification and replay (.) as essential for regaining bearings after becoming disoriented in complex data spaces.
  \item [DG4.]\textbf{\textit{ The prototype should incorporate stereo and volumetric spatial audio to encode position and depth cues:}} Stereo panning and distance-based loudness was collectively decided upon as the most effective method to convey horizontal and vertical positioning within surface plots.
  \item [DG5.]\textbf{\textit{ The prototype should implement configurable buffer playback to balance analytic detail and cognitive load:}} The ability to toggle between aggregated and non-aggregated modes allowed co-designers to select their customized set of values to be replayed and compared; giving them the ability to adjust buffer density depending on task complexity and preference.
  
\end{itemize}

\subsection{\textbf{Phase 2, Session 2: Testing and Validating New Features}}
\label{subsec:findings_phase2-session2}
Session 2 served two functions: validating the three features implemented after Session 1 and brainstorming future directions for embodied interfaces. The validation phase systematically tested each feature through task-specific prompts, while the brainstorming phase generated speculative designs for translating the prototype beyond web-based interaction, in line with embodied and immersive accessibility research~\cite{liu_interactive_2022, jamaludinAnsweringWhyWhen2023}.

\subsubsection{\textbf{Part 1: Testing of features}}
\label{subsubsec:findings_phase2-session2-part1}
This validation phase addressed \textbf{RQ2} by systematically testing whether each implemented feature successfully supported its target analytic tasks. Each feature was evaluated through specific task-based prompts, following task-centered evaluation practices in accessible visualization research~\cite{sharif_understanding_2021}: orientation restoration for \textit{DG3}, gradient tracing for \textit{DG4}, and region comparison for \textit{DG5}. Table~\ref{tab:session2-validation} summarizes the test prompts, task performance results, and refinements for each validated feature. The reference sonification and replay mechanism (\textit{DG3}) received immediate validation. Co-designers found it essential for restoring orientation after extended navigation. Sile's observation that the feature allowed both coincidental and sequential playback highlighted its flexibility; users could hear the reference and current position simultaneously or separately depending on their analytic need. Venkatesh's positive assessment confirmed that the feature successfully addressed the spatial disorientation identified in Session 1.

Spatial audio (\textit{DG4}) yielded mixed results. Stereo panning effectively encoded horizontal position (X-axis), with Sile noting its utility for tracing gradients. However, volume alone proved insufficient for depth perception (Z-axis). Sile could not perceive the volumetric changes during navigation. JooYoung's suggestion to augment volume with reverb emerged directly from this limitation. His proposal to combine amplitude decay with echo creates a multi-dimensional depth cue that reinforces distance perception through both loudness and spatial reflection. This refinement exemplifies the iterative validation process: co-designers not only identified failures but proposed empirically grounded solutions rooted in their perceptual expertise.

The configurable buffer (\textit{DG5}) validated as a powerful analytic concept but surfaced critical usability issues, consistent with prior findings that aggregation mechanisms require explicit feedback to support interpretation~\cite{lundgard_accessible_2021}. Venkatesh questioned the two-step selection initiation process, suggesting that the system should anchor the selection at the user's current position when D is pressed, rather than requiring a separate Enter confirmation. Ken's question about the aggregated sound's utility revealed that the feature's purpose was not immediately apparent. Venkatesh clarified the intended use case: comparing aggregated values to individual points within the selected region. This narrative exposed a gap in the interaction design; the feature lacked sufficient feedback to communicate its analytic function. Venkatesh further suggested that the system announce when users enter or exit the selected region during subsequent navigation, providing continuous spatial awareness of the buffer's boundaries. These findings further refined our understanding of \textbf{RQ2}, demonstrating that empowering accurate task performance and confident interpretation requires not only implementing features but ensuring their interaction design clearly communicates how they support specific analytic workflows.

\begin{table*}[h]
  \centering
  \scriptsize
  \begin{tabular}{p{2.5cm} p{3cm} p{3.5cm} p{4.5cm}}
    \toprule
    \textbf{Feature} & \textbf{Test Prompt} & \textbf{Result \& Co-designer Quote} & \textbf{Next Steps / Refinements} \\\\
    \midrule
    Reference Sonification \& Replay (\textit{DG3}) & ``Navigate away from the origin and use the reference key to reorient yourself.'' & \textbf{Success.} Sile: ``it does help, actually... having it separate like that, but allowing you to play either coincidentally or in sequence, that's really nice.'' Venkatesh: ``that was the main observation I had, but I think it's pretty cool that I get to hear the reference point.'' & None required; feature successfully validated for orientation restoration. \\\\
    \midrule
    Stereo \& Volumetric Audio (\textit{DG4}) & ``Trace a horizontal path and describe the spatial cues you perceive.'' & \textbf{Partial success.} Stereo panning validated. Sile: ``(stereo panning) it actually has been very interesting.'' Volume for depth too subtle. Sile: ``(volume) I couldn't notice it that much.'' & Add reverb to depth encoding. JooYoung: ``Using volume only may not be sufficient, so I suggest using echo sound (reverb)...if you combine volume plus, echo, then, it will be more pronounced.'' \\\\
    \midrule
    Configurable Buffer (\textit{DG5}) & ``Select a region, compare its aggregated sound to individual points, then compare the region to a different area.'' & \textbf{Partial success.} Concept validated, interaction issues surfaced. Venkatesh: ``why...I mean, just flip the order of actions, right? Like, first go to the point that you want to start selection from, and then when you press D, the backend knows.'' Ken: ``I'm wondering...How this is useful... what am I getting from that?'' & Redesign interaction: anchor selection at current position on first D-press. Add boundary entry/exit announcements. Venkatesh: ``my use case for suggesting this is comparing whatever you might hear in aggregate to individual points.'' See Table~\ref{tab:buffer-usability} for prioritized fixes. \\\\
    \bottomrule
  \end{tabular}
  \caption{Session 2 feature validation: test prompts, results, and refinements.}
  \label{tab:session2-validation}
  \Description{Table summarizing Session 2 validation results for three features. Reference sonification: Successfully validated for orientation restoration. Spatial audio: Stereo panning validated but volume insufficient for depth; refinement adds reverb per JooYoung's suggestion. Configurable buffer: Concept validated but interaction needs redesign; Venkatesh and Ken identify usability issues including two-step initiation and unclear aggregation purpose. Each row includes test prompts, co-designer quotes, and specific refinement recommendations.}
\end{table*}

Co-designers discovered they could select a region using the buffer, then invoke the AI assistant to ask targeted questions about only that subset of the data. This scoped query pattern aligns with recent work on coupling spatial selection with conversational analysis~\cite{reinders_when_2025}. This scoped query pattern emerged organically from Venkatesh's observation that ``if I'm looking at just this peak, I want to ask the AI about just this peak, not the whole dataset.'' The buffer mechanism, originally designed for audio-based region comparison, thus revealed a second analytic function: constraining the context window for natural language queries. This finding demonstrates how multimodal features can support complementary interaction modalities—spatial selection via keyboard navigation directly informs textual query scope, creating a tighter coupling between embodied exploration and AI-mediated analysis. Table~\ref{tab:buffer-usability} documents the prioritized issues surfaced during buffer interaction testing and the corresponding design refinements necessary to support both audio comparison and AI scoping workflows.

\begin{table*}[h]
  \centering
  \scriptsize
  \begin{tabular}{p{1cm} p{4cm} p{4.5cm} p{3.5cm}}
    \toprule
    \textbf{Priority} & \textbf{Usability Issue} & \textbf{Co-designer Quote / Observation} & \textbf{Proposed Fix} \\\\
    \midrule
    P1 & Two-step selection initiation breaks flow; users must navigate away, press D, navigate to start point, press Enter. & Venkatesh: ``why...I mean, just flip the order of actions, right? Like, first go to the point that you want to start selection from, and then when you press D, the backend knows.'' & Anchor selection at current position on first D-press; eliminate redundant Enter step. \\\\
    \midrule
    P2 & Lack of boundary awareness; users cannot tell when they re-enter or exit the selected region during subsequent navigation. & Venkatesh: ``it would be nice if, as you're navigating around...it announces when you enter the region or when you leave the region.'' & Add audio announcement when cursor crosses buffer boundary; consider persistent background cue (e.g., subtle ambient tone) while inside region. \\\\
    \midrule
    P3 & Aggregated playback purpose unclear; co-designers unsure when mean value is useful versus sequential point-by-point replay. & Ken: ``I'm wondering...How this is useful... what am I getting from that?'' Venkatesh clarified: ``my use case for suggesting this is comparing whatever you might hear in aggregate to individual points.'' & Provide contextual help (e.g., on-demand description of aggregation modes); add example prompts in tutorial demonstrating comparison workflow. \\\\
    \midrule
    P4 & No visual or audio confirmation when buffer selection is saved; uncertain whether system registered the region. & Observed during testing: co-designers pressed G multiple times, unsure if buffer was stored. & Add distinct confirmation sound when buffer is saved; optionally announce region dimensions (e.g., ``Buffer saved: 6 by 7 region''). \\\\
    \bottomrule
  \end{tabular}
  \caption{Prioritized buffer usability issues and proposed fixes.}
  \label{tab:buffer-usability}
  \Description{Priority-ranked table of four buffer usability issues with proposed solutions. P1: Two-step selection breaks flow; fix by anchoring at current position. P2: No boundary awareness; add entry/exit announcements. P3: Unclear aggregation purpose; provide contextual help and examples. P4: No save confirmation; add confirmation sound and region dimensions announcement. Each issue includes supporting co-designer quotes from Venkatesh and Ken.}
\end{table*}

\subsubsection{\textbf{Part 2: Discussed Future Directions}}
\label{subsubsec:findings_phase2-session2-part2}
The brainwriting portion of \textbf{Session 2} was \textbf{\textit{primarily focussed on answering RQ3}}, generating a rich set of ideas for moving beyond the current web-based implementation. These ideas coalesced around three interconnected themes that align with prior embodied and immersive accessibility research~\cite{liu_research_2020, johnsonSupportingDataVisualization2023}: advanced haptics to restore persistent tactile feedback, mixed and augmented reality to merge physical and digital exploration spaces, and alternative physical interfaces to replace keyboard-centric navigation. Collectively, these proposals represent co-designers' vision for translating the intuitive, exploratory nature of the tactile probe into digitally mediated forms that preserve, and potentially augment, its embodied affordances.

\begin{enumerate}
    \item \textbf{Advanced Haptics and Tactile Displays:} Haptic feedback emerged as the most direct pathway for restoring the persistent tactile cues that grounded the physical prototype. Sile articulated haptics' strength: ``really good at dynamic feedback,'' proposing that force resistance could encode data density—an aggregated data point ``could be heavier, or harder to press against, the more data it is aggregating.'' This maps physical effort directly onto information density, creating an embodied metric for data complexity. Ken identified near-term implementation targets, advocating for existing devices like the Inverse3 Haptic Controller\footnote{https://www.haply.co/inverse3} or the Graphiti tactile display\footnote{https://www.orbitresearch.com/product/graphiti-plus} as a ``quickstep to trying a different method of 'feeling' the graph.'' Sile extended this vision to multi-line braille displays, imagining parallel multi-finger sensing that would allow simultaneous selection and comparison of multiple regions—a shift from sequential keyboard navigation to parallel tactile interaction. These proposals share a common thread: leveraging haptic modalities to restore the simultaneity and persistence lost in the transition from physical to digital.
    \item \textbf{Mixed and Augmented Reality:} MR aims to merge the physical exploration space with digital data overlays. JooYoung positioned MR as ``critical,'' envisioning ``bidirectional synchronization between the digital environments and physical artifacts.'' His three-step operationalization concretizes this vision: (a) construct a physical play mat demarcating the X-Z grid, (b) align the AR boundary with the mat's physical edges, and (c) sonify Y-values as pitch while users physically traverse the mat. This design recasts the user's body as an ``embodied cursor,'' directly mapping physical movement through space onto data navigation. The approach dissolves the screen boundary, allowing co-designers to ``walk through'' the dataset using proprioceptive and vestibular cues—modalities entirely absent from keyboard-based interaction. This proposal directly extends the tactile prototype's affordances by preserving spatial exploration while augmenting it with dynamic, real-time sonification.
    \item \textbf{Alternative Physical Interfaces:} Beyond haptics and MR, co-designers proposed novel input devices to escape keyboard constraints while remaining grounded in accessible hardware. Venkatesh suggested repurposing familiar controls: two-way trackballs, surface dials, laptop touchpads, or even the keyboard surface itself as a 2D positional sensor. These ideas prioritize learnability and availability, adapting existing assistive technology interaction patterns to 3D data navigation. Ken proposed a more elaborate system: a glove-mounted hand tracker controlled by CNC-style XYZ pulleys, providing kinesthetic feedback as users ``sculpt'' through the data space. This design borrows from haptic sculpture interfaces, mapping fine motor control onto precise data interrogation. Collectively, these speculative interfaces share a rejection of traditional keyboard navigation in favor of controls that more closely mirror the physical prototype's directness—whether through rotational input, positional sensing, or kinesthetic manipulation. Each proposal answers RQ3 by translating a specific dimension of the tactile probe's embodied affordances into digitally augmented form.
\end{enumerate}
\section{Discussion}
\label{sec:discussion}
Moving beyond a simple summary of outcomes, this section reflects on the insights from our iterative process. We interpret our findings by connecting them to existing academic literature and systems, focusing on the conceptual challenges of knowledge transfer from physical to digital forms, the nuanced role of non-visual features in empowering users, and the emerging pathways toward more physically-grounded data interaction~\cite{marriott_inclusive_2021, elavsky_how_2022, jamaludin_answering_2023}.

\subsection{Limitations of Our Study}
\label{subsec:limitations}
Our study's methodological scope introduces five primary limitations that constrain the generalizability of our findings. First, \textbf{expert sample bias}: our six BLV co-designers all possess advanced technical literacy, extensive experience with assistive technologies, and professional backgrounds in data analysis or accessibility research. While this auto-ethnographic approach provided invaluable, high-fidelity insights grounded in lived expertise, features that were intuitive for this group may not be learnable for novice users unfamiliar with data visualization or advanced assistive technologies~\cite{sharif_understanding_2021, zong_rich_2022}. Our prototype reflects design priorities and interaction patterns that emerged from expert users who bring sophisticated mental models and analytic strategies to 3D data exploration, which may not align with the needs, preferences, or learning trajectories of BLV individuals with less prior exposure to data visualization, scientific computing, or spatial reasoning tasks. Second, \textbf{limited dataset types}: our evaluation was conducted using relatively clean, dense datasets (benzene spectroscopy, Gaussian surfaces, weather data). The effectiveness of our sonification and navigation strategies has not been tested on sparse, disjointed, or noisy datasets, where features like peak-jumping or aggregated buffering might behave differently~\cite{lapusteanu_review_2024}. Third, \textbf{cross-sectional study design}: we captured co-designers' immediate reactions during two sessions but have not assessed longitudinal effects of using the prototype over time, where issues like auditory fatigue, habituation, or the development of new usage patterns might emerge~\cite{enge_open_2024}. Fourth, \textbf{volumetric audio subtlety}: as identified in Session 2, volume alone proved insufficient for depth perception (Z-axis), requiring augmentation with reverb—a refinement not yet implemented in the validated prototype~\cite{siu_virtual_2020}. Fifth, \textbf{buffer UX issues}: the configurable buffer surfaced critical usability barriers documented in Table~\ref{tab:buffer-usability}, including two-step selection initiation, lack of boundary awareness, unclear aggregation purpose, and absent confirmation feedback. While the buffer concept validated successfully, its interaction design requires substantial refinement before deployment with non-expert users.

\subsection{Conceptual Contributions Derived from the Co-Design Study}
\label{subsec:conceptual-contributions}
Our Experience-Based Co-Design process with expert BLV collaborators yielded three interrelated conceptual contributions that extend beyond the specific implementation of our system. These contributions address longstanding gaps in accessible 3D data visualization by centering lived experience as design material~\cite{raynor_experiencebased_2020, morley_evidence-informed_2024, lundgard_sociotechnical_2019}.

\begin{itemize}
    \item We developed a \textbf{transferable co-design protocol for translating tactile expertise into multimodal digital interaction}. By pairing a low-fidelity tactile probe with a high-fidelity digital prototype, we established a methodological framework that systematically elicits tactile knowledge~\cite{vandemoerePhysicalVisualizationInformation2010} and translates it into non-visual digital affordances. This protocol operationalizes EBCD principles~\cite{raynor_experiencebased_2020} for HCI contexts where physical and digital modalities must interoperate. The tactile probe served as an epistemological anchor, providing a shared sensory baseline against which digital experiences could be evaluated. Co-designers' narratives about orientation confusion, spatial comparison challenges, and analytic workflows directly motivated implemented features (reference sonification, configurable buffer, spatial audio), demonstrating how experiential ``touch-points''~\cite{graber_reflections_2024} can be systematically converted into design requirements. This protocol is transferable: future projects targeting other 3D data types (e.g., molecular structures, terrain models) can adapt our phased approach; tactile grounding, comparative evaluation, iterative refinement, to surface domain-specific analytic needs.
    \item We produced a \textbf{validated prototype enabling independent non-visual 3D data exploration}. Our system demonstrates that web-native, multimodal interaction can support core analytic tasks (3D orientation, gradient tracing, landmark identification, region comparison) without requiring specialized hardware. \textbf{Session 2} validation confirmed that features like reference sonification, stereo panning, and buffer aggregation improved co-designers' ability to orient, analyze, and interpret 3D surfaces with confidence. Unlike prior systems that rely on expensive tactile displays~\cite{suzuki_fluxmarker_2017, leithinger_shape_2015} or sighted-centered VR~\cite{jamaludin_answering_2023}, our web-based approach offers scalable, low-cost access to continuous height-field surfaces across scientific domains. The prototype's modular architecture enabled rapid iteration during co-design sessions, allowing us to implement complex features (e.g., buffer aggregation with dual playback modes) within days of co-designer feedback. This responsiveness exemplifies how technical infrastructure can support rather than constrain participatory design~\cite{sanders_co-creation_2008}.
    \item We established \textbf{empirically grounded design principles for accessible visualization}. Our findings converge on four principles that extend existing accessible visualization guidance~\cite{marriott_inclusive_2021, elavsky_how_2022} into three-dimensional contexts: (1) \textit{Multi-parameter redundancy}; (2) \textit{Progressive disclosure}; (3) \textit{Buffer-aggregation as analytic scoping}; and (4) \textit{Persistent spatial reference}. These principles, validated through co-designer testimony and artifact evaluation, offer concrete guidance for designers extending accessible visualization beyond 2D charts into volumetric, spatially continuous datasets~\cite{enge_open_2024}.
\end{itemize}

\subsection{System-based Design Insights}
\label{subsec:system-insights}
Drawing from both our prototype implementation and co-designer validation, we distill two complementary sets of insights: boundary conditions specific to our system's current instantiation, and design principles applicable to future accessible 3D visualization systems. This dual framing, requested by co-designers during Session 2 debriefing, clarifies which design decisions are implementation-dependent versus conceptually transferable~\cite{huang_experiential_2023}.

\subsubsection{Boundary Conditions and Prototype Specifics}
\label{subsubsec:boundary-conditions}
Our system's validated configuration represents a constellation of design decisions that emerged iteratively through co-designer feedback and artifact evaluation. We document these boundary conditions to distinguish implementation-specific choices from transferable principles, clarifying the scope within which our findings remain empirically grounded.

\begin{itemize}
    \item \textbf{Binaural audio engine}: Our browser-native implementation is built entirely on Web Audio API primitives operating at 48kHz sample rate, deliberately avoiding external synthesis libraries to ensure zero-latency compatibility across devices. The audio signal chain progresses sequentially from oscillator generation through envelope shaping, stereo panning, and convolution reverb, creating a modular architecture that our co-designers found responsive during real-time exploration. This architectural choice prioritized immediate auditory feedback over sophisticated synthesis capabilities, which was validated when co-designers consistently praised the system's responsiveness during navigation.
    \item \textbf{Sonification mappings}: Spatial coordinates translated into perceptible audio through four concurrent encodings that emerged directly from Session 1 deliberations. Following Aziz's suggestion, horizontal position (X-axis) maps to stereo panning ranging from fully left (-1.0) to fully right (+1.0), leveraging users' innate spatial hearing capabilities. Vertical position (Y-axis, representing surface height) maps to pitch via a logarithmic scale spanning 200Hz to 800Hz, ensuring that equal perceptual differences correspond to proportional height changes~\cite{hoque_accessible_2023}. Depth perception (Z-axis) proved most complex, ultimately requiring JooYoung's proposed multi-parameter encoding that combines dynamic volume with frequency-dependent reverb: closer points play at higher amplitude (20\% wet mix) with minimal pre-delay (10ms) and preserved high frequencies (6500Hz lowpass), while distant points grow quieter (95\% wet mix), gain substantial echo (90ms pre-delay), and lose treble clarity (2000Hz lowpass). Additionally, waveform timbre varies—sine waves for low-frequency regions, triangle waves for mid-range, square waves for high-frequency areas—to provide redundant height information through timbral brightness.
    \item \textbf{Configurable buffer mechanism}: JooYoung's Session 1 brainwriting proposal established rectangular region selection through a three-stage keyboard interaction: D-key initiates selection mode, Enter anchors the starting corner at the current cursor position, arrow keys navigate to define the opposite corner, and a second Enter press confirms the bounded region. The buffer stores all grid rectangles or individual data points falling within these bounds, enabling subsequent analytic operations. When users activate aggregation mode via G-key, the system computes the arithmetic mean of normalized Y-values within the buffered region and plays a sustained 1.0-second tone representing this summary statistic. Alternatively, sequential playback mode traverses each buffered item individually at 0.3-second intervals with 125ms inter-item silence, allowing detailed inspection. These timing parameters emerged through Aziz's iterative testing during artifact evaluation, balancing playback duration against cognitive processing demands.
    \item \textbf{Dataset density constraints}: Our validation focused exclusively on dense, continuous surface datasets where neighboring points maintain close spatial proximity: benzene VUV spectroscopy data (3,116 measurement points), synthetic Gaussian surfaces (2,500 points), and meteorological height fields (1,200 points). These dataset constraints have direct implications: peak-jumping functionality reliably identifies local maxima when surfaces exhibit smooth gradients, but its behavior remains undefined for sparse point clouds or disjoint data distributions with large spatial voids. Similarly, buffer aggregation produces meaningful summary statistics when regions contain sufficient data density to yield representative means, a condition not guaranteed for irregular sampling patterns. We explicitly acknowledge this limitation in Section~\ref{subsec:limitations}, recognizing that alternative dataset types may require modified algorithms or additional safeguards. 
    \item \textbf{Two-tier navigation structure}: Navigation defaults to a 20$\times$20 wireframe grid overlay where each rectangular cell aggregates the underlying surface region it bounds, reducing the initial exploration space from potentially thousands of individual points to 400 manageable zones. Aziz specifically requested this ``surface mode'' during Phase 1 co-design to mitigate cognitive load when first encountering unfamiliar datasets, allowing users to grasp overall topography before diving into granular detail. Users can toggle to point mode via 2-key when finer-grained analysis becomes necessary, shifting from regional overviews to vertex-by-vertex traversal. This structure mirrors cartographic conventions of zooming from overview maps to detailed street views~\cite{roth_cartographic_2021}.  
    \item \textbf{Spatial orientation support}: Two complementary mechanisms that Ken and Sile advocated during Session 1 was about reference sonification. It provides an absolute coordinate anchor: pressing 0-key triggers a fixed 300Hz sine tone representing the dataset origin (0,0,0) regardless of current cursor position, allowing users to recalibrate their mental spatial model at any moment. Complementing this, the replay mechanism (. key) re-sonifies the user's current position without causing navigation, enabling repeated listening when interpreting ambiguous or complex local features. These orientation aids address the fundamental challenge of maintaining spatial awareness in non-visual environments where users cannot glance at coordinate axes as sighted analysts routinely do.
\end{itemize}

These specifications collectively define the empirical scope of our co-design validation. We present them not as prescriptive requirements but as documented implementation choices that proved effective with our expert co-designer cohort exploring dense surface datasets via keyboard interaction. Different user populations, alternative dataset types, or novel deployment contexts may necessitate adjusted parameter ranges, modified interaction patterns, or architectural alternatives.

\subsubsection{Design Principles}
\label{subsubsec:design-principles}
Beyond our system's specific configuration, our co-design process surfaced four transferable principles applicable to accessible 3D visualization systems regardless of implementation platform.

\begin{itemize}
    \item \textbf{Multi-parameter redundancy for spatial encoding}: Encoding each spatial dimension through multiple reinforcing modalities (e.g., X-axis via stereo panning \textit{and} verbal coordinate announcements; Z-axis via volume \textit{and} reverb \textit{and} textual depth cues) ensures robust perception across diverse sensory preferences and abilities. Co-designers with varying visual acuities reported differential reliance on auditory vs. textual feedback, validating the necessity of parallel channels~\cite{hoque_accessible_2023, jiang_designing_2024}. \textit{Source: Co-designer testimony (Session 2); Sile noted reliance on text, Venkatesh prioritized audio.}
    \item \textbf{Progressive disclosure through whole-to-part navigation}: Scaffolding users from high-level overviews (auto-play across multiple perspectives) to granular detail (point-by-point traversal) supports pedagogical transitions from initial orientation to hypothesis-driven analysis. This mirrors established accessible visualization patterns~\cite{seo_maidr_2024, zhang_charta11y_2024} but extends them into volumetric contexts. The ability to quickly ``snapshot'' a dataset's topography before diving into specific features reduces cognitive load and accelerates mental model formation. \textit{Source: Co-designer testimony (Phase 1); JooYoung identified tedium of point-by-point exploration.}
    \item \textbf{Buffer-aggregation as analytic scoping tool}: Enabling users to define spatial regions of interest and toggle between summary statistics (mean) and sequential detail empowers two critical workflows: (1) comparing local maxima to surrounding context, and (2) scoping natural language queries to specific subsets of data. This principle extends beyond sonification; any multimodal system benefits from user-controlled granularity adjustments. The emergent use case of buffer-scoped AI queries (Section~\ref{subsec:findings_phase2-session2}) demonstrates how spatial selection mechanisms can integrate with conversational interfaces. \textit{Source: Co-designer testimony (Session 2); Venkatesh articulated comparison use case, emergent AI scoping observed during testing.}
    \item \textbf{Pedagogical scaffolding for tactile-to-digital transition}: Designing digital features to replicate the cognitive functions (not literal affordances) of physical probes facilitates knowledge transfer. Our tactile probe's persistent frame of reference became digital reference sonification; its multi-finger comparison became configurable buffer playback. This principle aligns with mixed-reality prototyping best practices~\cite{snider_mixed_2024, speicher_designers_2021} but specifies a pathway for non-visual contexts: deconstruct physical affordances into core cognitive operations, then implement digital analogs serving those operations. \textit{Source: Co-designer testimony (Session 1 comparative analysis); conceptual mapping documented in Section~\ref{subsec:findings_phase2-session1}.}
\end{itemize}

The principles aforementioned, grounded in co-designer narratives and validated through artifact evaluation, provide actionable guidance for future accessible 3D visualization systems. Unlike the boundary conditions in Section~\ref{subsubsec:boundary-conditions}, these insights transcend specific implementation choices and apply across platforms, modalities, and dataset types.

\subsection{Exploratory Insights from Co-Designer Speculations}
\label{subsec:exploratory-insights}
Session 2’s brainwriting exercise (Section~\ref{subsubsec:findings_phase2-session2-part2}) produced speculative extensions of the validated web-based prototype toward more embodied forms of 3D data exploration. These concepts were not implemented or evaluated; instead, they reflect co-designers’ expert intuitions about how embodied cognition~\cite{liu_research_2020} and proprioception~\cite{ball_realizing_2007} could support non-visual analysis beyond screen-based interaction. We present these ideas as exploratory insights that outline possible future directions rather than validated design outcomes.

\subsubsection{Advanced Haptics as Persistent Tactile Grounding}
\label{subsubsec:advanced-haptics}
Co-designers consistently identified haptic feedback as the most promising mechanism for restoring a sense of physical persistence to data exploration. Mechanisms to encode data density through force resistance, allowing users to ``feel'' aggregated regions as heavier or harder to press against; maps analytic properties (e.g., density, gradients) onto physical effort, providing continuous, user-controlled cues rather than transient auditory signals. Co-designers also proposed large, multi-line braille displays to enable parallel, multi-finger comparison across regions of a dataset. These ideas build on prior work in touch-based accessible graphics~\cite{butlerTechnologyDevelopmentsTouchBased2021} and shape-changing interfaces~\cite{leithinger_shape_2015}, but emphasize haptics as an analytic encoding rather than a visual surrogate. Future work could prototype task-specific haptic mappings (e.g., force gradients for slope, vibration for density) and compare their learnability to sonification. \textit{Source: Co-designer speculation (Session 2 brainwriting).}

\subsubsection{Embodied Spatial Exploration Beyond the Screen}
\label{subsubsec:embodied-exploration}
Mixed and augmented reality were discussed only as background inspiration for extending interaction beyond the screen, rather than as implementation targets. Co-designers emphasized the value of leveraging whole-body proprioception, such as walking or reaching, to support spatial reasoning in 3D data. Related suggestions focused on spatialized audio, with data points rendered as a 3D soundscape that users could explore by moving through space. These ideas align with prior work on immersive and embodied analytics~\cite{liu_interactive_2022, jamaludin_answering_2023}, but were not central to the prototype’s design or evaluation. Instead, they highlight longer-term opportunities for coupling proprioception and spatial audio to support non-visual mental models of complex data. \textit{Source: Co-designer speculation (Session 2 brainwriting).}

\subsubsection{Alternative Physical Interfaces Beyond Keyboard Navigation}
\label{subsubsec:alternative-interfaces}
Co-designers also critiqued the mismatch between discrete keyboard input and continuous 3D data spaces~\cite{wang_web-based_2005}. Arrow-key navigation enforces sequential, axis-aligned movement, increasing cognitive load when exploring continuous surfaces. Across proposals (Section~\ref{subsubsec:alternative-interfaces}), a shared principle emerged: \textit{input modalities should match the dimensionality and continuity of the data}. Future work could compare navigation efficiency, learnability, and analytic accuracy across keyboards, analog controllers, and haptic devices, particularly for users with diverse motor abilities and assistive technology experience. \textit{Source: Co-designer speculation (Session 2 brainwriting).}

\subsection{Future Directions}
\label{subsec:future-direction}
Our future work will move from foundational design to rigorous validation and conceptual expansion. The immediate priority is to conduct formal user studies to validate the usability and learnability of our current feature set with a broader, non-expert BLV population. Following this, we will pursue a research agenda focused on enhancing the user's control and perception. This includes investigating alternative physical input devices, such as joysticks and trackballs, to afford more fluid, multi-dimensional navigation that better aligns with the exploration of 3D space. We will also refine the sonification engine, exploring more perceptually salient audio cues, such as reverb, to create a richer and more intuitive representation of depth. Finally, we will expand the prototype’s analytical scope by implementing features for more complex data scenarios, such as the comparative plot view, to study how non-visual users perform analysis across layered datasets. We will also attempt to prototype a spatially adaptive interface to examine how proprioception can transform data exploration from a navigational task into an immersive experience. 

\section{Conclusion}
\label{sec:conclusion}


Through an experience-based co-design process with BLV collaborators with expertise in non-visual data representations, our aim was to create a digital experience that preserves spatial understanding and enables independent multiperspectival exploration that is scalable into multiple fields irrespective of data types. Our refined prototype demonstrated how sonification, spatialized audio, and adaptive aggregation can support orientation, accuracy, and learnability in surface and point plots. Beyond the artifact, our work contributes (1) a transferable co-design protocol for translating tactile expertise into digital interaction, and (2) concrete guidance for designing accessible analytic workflows around volumetric data.
At the same time, our findings are bounded by a small expert user group, a limited set of datasets, and a cross-sectional evaluation. Future work should expand evaluation with a broader BLV population, investigate alternative input modalities for more fluid navigation, refine sonification with perceptually salient cues, and extend analytic scope to comparative and embodied interfaces. Taken together, these directions chart a concrete, evidence-based path toward equitable access to 3D scientific visualization.

\begin{acks}
  This work was supported in part by JooYoung Seo's Faculty Startup Fund, which provided funding for project materials, including low-fidelity prototypes and shipping costs. We would also like to extend our sincere gratitude to Tad Schroeder, an overall superhero, day-jobbing as the Assistant Director of Facilities at the School of Information Sciences, University of Illinois
Urbana-Champaign for his invaluable support in coordinating the shipping logistics for our low-fidelity prototype.
\end{acks}

\bibliographystyle{ACM-Reference-Format}
\bibliography{references/references, references/extra, references/related}

\appendix

\section{Brainwriting Session Prompts and Task-Based Questions}

\subsection{Session 1: Brainwriting Prompts}
\label{subsec-app:session1-bwprompt}
\begin{enumerate}
    \item \textbf{\textit{Which elements of the tactile experience feel essential for understanding the surface?}} This question was designed to focus the team on the core affordances of the physical prototype, identifying what made it effective for building a spatial mental model.
    \item \textbf{\textit{How might these translate into a digital interface?}} This prompts a direct translation of the previously identified essential traces of the physical prototype into potential digital features, bridging the gap between the physical and virtual.
    \item \textbf{\textit{What features or feedback would help reproduce the tactile clarity in a web interface?}}The aim was to focus on the specific quality of ``clarity'', encouraging co-designers to think about the fidelity and type of feedback needed to make the digital experience as unambiguous as the physical one.
    \item \textit{\textbf{What risks do we face if these features aren't included?}} This prompt helped us consider what would be lost in a poor translation. Our aim was to brainstorm the most critical design requirements by forcing a consideration of failure modes.
\end{enumerate}

\subsection{Session 1: Extended Task-Based Questions}
\label{subsec-app:session1-tasks}
Throughout Session 1's probe exploration and brainwriting exercise, co-designers engaged with the following task-based questions, which were presented conversationally to elicit their exploration strategies, confusions, and insights:

\subsubsection*{3D Orientation Tasks}
\begin{itemize}
    \item How do you know where you are in the plot when using the tactile probe versus the digital prototype?
    \item What cues help you understand the axes in each version?
    \item When you lose your orientation, what helps you recover your bearings?
    \item How would you differentiate between the X, Y, and Z axes nonvisually?
\end{itemize}

\subsubsection*{Landmark and Peak Finding Tasks}
\begin{itemize}
    \item How would you locate the highest point on the surface using the tactile probe?
    \item How would you match this and locate the highest point using the digital prototype?
    \item Can you find where the data changes most dramatically? What strategies do you use?
    \item What features or feedback would help you quickly jump to salient points of interest?
\end{itemize}

\subsubsection*{Comparing Local Maxima versus Global Trends}
\begin{itemize}
    \item How do you differentiate a single peak from an overall rising trend in the tactile probe?
    \item What would help you compare multiple high points in the digital interface (key-binding or continuous sonification)?
    \item When exploring the surface, how do you maintain awareness of the overall shape while examining local features?
    \item What feedback mechanisms would support comparison between different regions?
\end{itemize}

\subsubsection*{Gradient Tracing Tasks}
\begin{itemize}
    \item Can you follow the path of steepest ascent on the physical probe? How?
    \item How would you trace a ridge or valley in the digital version?
    \item What auditory or tactile cues would best indicate the direction and steepness of slopes?
    \item How do you distinguish between gradual slopes and sharp cliffs?
\end{itemize}

\subsubsection*{Identifying Occluded or Partially Hidden Features (for those with Residual Vision)}
\begin{itemize}
    \item How do you explore areas that might be hidden from one perspective on the tactile probe?
    \item What strategies help you uncover features that aren't immediately apparent in the digital interface?
    \item How would you know if there are features "behind" or "underneath" other parts of the surface?
    \item What feedback would indicate overlapping or complex spatial relationships?
\end{itemize}

\subsection{Session 2: Brainwriting Prompts}
\label{subsec-app:session2-bwprompt}
\begin{enumerate}
    \item \textbf{\textit{Which tools, techniques, or mediums can help preserve embodied cognition digitally?}} This question was intentionally broad, using the term ``embodied cognition'' to encourage the team to think beyond simple keyboard interactions and consider how physical movement and presence could be integrated.
    \item \textbf{\textit{How can sound, text or haptics replicate the spatial clarity of the physical prototype - shifting from a web-based prototype to a spatially adaptive interface to support embodied cognition?}} This prompt focuses on the specific modalities (sound, text, haptics) and the goal of creating a ``spatially adaptive interface'', pushing the team to ideate on how the system could respond to a user's physical context.
    \item \textbf{\textit{What's the most feasible and understandable step to move toward these interfaces?}} After the expansive ideation prompted by the first two questions, the idea was to ground the discussion in pragmatism, asking the team to identify a concrete, actionable first step toward their more ambitious, long-term visions.
\end{enumerate}

\subsection{Session 2: Extended Task-Based Questions}
\label{subsec-app:session2-tasks}
During Session 2's feature validation phase, co-designers were asked to evaluate the newly implemented features through the following task-based questions:

\subsubsection*{Reference Sonification and Orientation (DG3)}
\begin{itemize}
    \item Does the fixed origin sound help you maintain orientation during exploration?
    \item How useful is the replay mechanism when you become disoriented?
    \item Can you use the reference point to navigate back to previously explored regions?
    \item What additional reference points or landmarks would be helpful?
\end{itemize}

\subsubsection*{Spatial Audio for Position and Depth (DG4)}
\begin{itemize}
    \item How effectively does stereo panning convey horizontal position (X-axis)?
    \item Can you perceive depth (Z-axis) through volume changes alone?
    \item What makes spatial audio cues more or less salient during exploration?
    \item How would additional audio cues (e.g., reverb, echo) enhance depth perception?
\end{itemize}

\subsubsection*{Configurable Buffer for Comparison (DG5)}
\begin{itemize}
    \item How intuitive is the buffer selection process?
    \item Does the aggregated playback help you understand regional patterns?
    \item Can you effectively compare a saved buffer region to your current focus point?
    \item What improvements would make the buffer more useful for analytical tasks?
    \item How might the buffer support focused queries to the AI assistant?
\end{itemize}

\subsubsection*{Multi-Perspective Auto-Play and Jump-to-Peak (DG1, DG2)}
\begin{itemize}
    \item How do the auto-play overviews help you build an initial mental model?
    \item Does jump-to-peak functionality help you efficiently locate salient features?
    \item What additional automated exploration strategies would be valuable?
    \item How do these features balance efficiency with user control?
\end{itemize}

\subsubsection*{Future Embodied Interface Concepts}
\begin{itemize}
    \item What physical input devices would best support 3D navigation?
    \item How could haptic feedback enhance your understanding of data density or slope?
    \item In a mixed reality environment, how would you use your body to explore the data space?
    \item What's the most important first step toward creating a spatially adaptive interface?
\end{itemize}

\section{AI Chat Assistant Implementation}
\label{subsec-app:ai-chat}
\begin{figure}[h!]
    \centering
    \includegraphics[width=\linewidth]{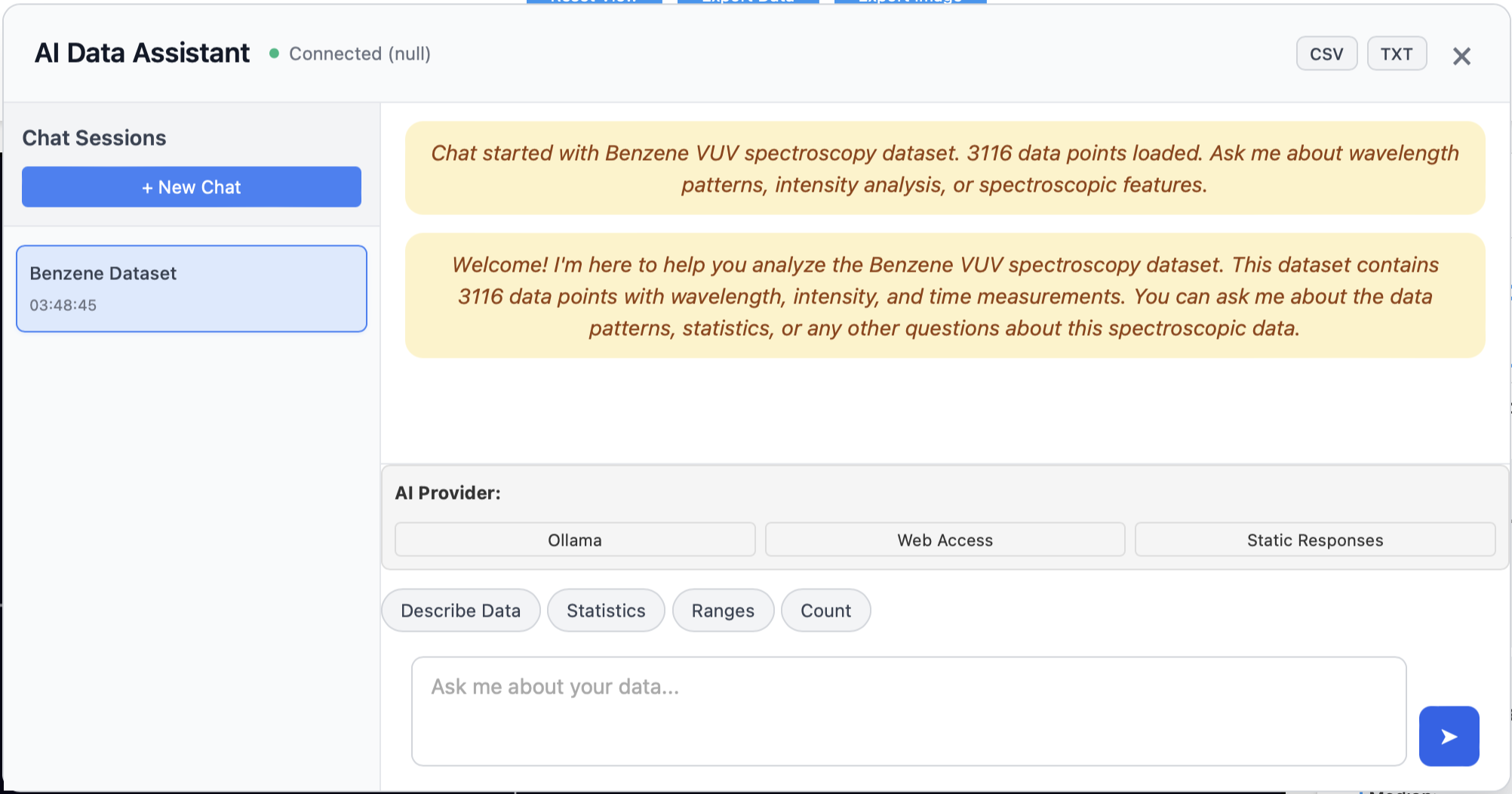}
    \caption{AI Chat Interface}
    \Description{AI Data Assistant interface showing active Benzene Dataset chat session. Left panel: Chat title, timestamp, and "+ New Chat" button. Center: Two yellow system messages welcoming user and confirming 3116 data points loaded for wavelength, intensity, and time analysis. Bottom: Query input bar with send button, three AI provider options (Ollama, Web Access, Static Responses), and four quick-action buttons (Describe Data, Statistics, Ranges, Count). Top right: CSV and TXT export options.}
    \label{fig:ai-interface}
\end{figure}

The multi-threaded AI chat assistant (viewable in Figure~\ref{fig:ai-interface}) leverages Gemini 2.5 Pro to support interactive data exploration. Since the WebGL plot supports cursor-based interactivity, a screenshot of the user's current view was passed to the model alongside the raw dataset used by the \emph{VisualizationEngine.js} from the \emph{Engine Layer}. This allows the model to provide contextually relevant explanations or respond to targeted questions about specific data points and regions. To ensure robustness, a system of static responses was also implemented, enabling the assistant to provide fallback answers based on pre-computed dataset statistics if the live model connection fails. For users with access to the system repository, if they chose to run it locally rather than through the web, the tool supported integration with Ollama, allowing queries to be directed to a user's local language model, including fine-tuned versions. The assistant was designed to automatically detect the most recent available model and establish a connection to facilitate seamless responses. Additionally, four example prompts were provided to guide users in framing queries to the assistant.

\end{document}